\documentclass[reprint,
 superscriptaddress,
nofootinbib,
 amsmath,amssymb,
 aps,
pra,
]{revtex4-2}

\usepackage[utf8]{inputenc} 
\usepackage[T1]{fontenc}    
\usepackage{hyperref}       
\usepackage{url}            
\usepackage{multirow}
\usepackage{booktabs}       
\usepackage{amsfonts}       
\usepackage{nicefrac}       
\usepackage{microtype}      
\usepackage{graphicx}
\usepackage{xcolor, etoolbox}
\usepackage{amsmath,amsfonts,amsthm} 
\usepackage{mathtools}
\usepackage{ragged2e}
\usepackage{braket}      
\usepackage[english]{babel} 
\usepackage{enumitem}
\usepackage{xspace}
\usepackage{dsfont}
\usepackage{bm}
\usepackage{bbm}
\usepackage{dcolumn}
\usepackage{mathrsfs}
\usepackage{fixltx2e}
\usepackage{comment}

\hypersetup{
    colorlinks,
    linkcolor=blue,
    citecolor=blue,
    urlcolor=blue,
    breaklinks=true 
}

\DeclarePairedDelimiter\abs{\lvert}{\rvert}

\DeclarePairedDelimiter\floor{\lfloor}{\rfloor}

\begin{document}

\title{From bosons and fermions to spins: \\ A multi-mode extension of the Jordan-Schwinger map}

\author{Benoît Dubus}
    \email{benoit.dubus@ulb.be}
    \affiliation{Centre for Quantum Information and Communication, École polytechnique de Bruxelles, CP 165, Université libre de Bruxelles, 1050 Brussels, Belgium}

\author{Tobias Haas}
    \email{tobias.haas@ulb.be}
    \affiliation{Centre for Quantum Information and Communication, École polytechnique de Bruxelles, CP 165, Université libre de Bruxelles, 1050 Brussels, Belgium}
    
\author{Nicolas J. Cerf}
    \email{nicolas.cerf@ulb.be}
    \affiliation{Centre for Quantum Information and Communication, École polytechnique de Bruxelles, CP 165, Université libre de Bruxelles, 1050 Brussels, Belgium}

\begin{abstract}
The Jordan-Schwinger map is widely employed to switch between bosonic or fermionic mode operators and spin observables, with numerous applications ranging from quantum field theories of magnetism and ultracold quantum gases to quantum optics. While the construction of observables obeying the algebra of spin operators across multiple modes is straightforward, a mapping between bosonic or fermionic Fock states and spin states has remained elusive beyond the two-mode case. Here, we generalize the Jordan-Schwinger map by algorithmically constructing complete sets of spin states over several bosonic or fermionic modes, allowing one to describe arbitrary multi-mode systems faithfully in terms of spins. As a byproduct, we uncover a deep link between the degeneracy of multi-mode spin states in the bosonic case and Gaussian polynomials. We demonstrate the feasibility of our approach by deriving explicit relations between arbitrary three-mode Fock and spin states, which provide novel interpretations of the genuinely tripartite entangled GHZ and W state classes.
\end{abstract}

\maketitle

\section{Introduction}
Utilizing mappings between observables representing \textit{different} types of degrees of freedom has a long tradition in solving complex quantum many-body models \cite{Klein1991}. Among many others, this includes the Jordan-Wigner map \cite{Jordan1928}, which relates interacting spin chains such as the Ising model with free fermions \cite{Lieb1961}, the Holstein–Primakoff map \cite{Holstein1940}, which maps spins to bosonic modes using expansions of the operator square root \cite{Vogl2020,Koenig2021}, as well as the Jordan-Schwinger map \cite{Jordan1935,Schwinger1952}, which establishes a direct mapping between Lie algebra representations and bosonic or fermionic modes.

Initially employed to characterize a single spin, more precisely, representations of the $\mathfrak{su}(2)$ algebra, in terms of two bosonic modes \cite{Schwinger1952,Sakurai2020}, the Jordan-Schwinger map found diverse applications in, for instance, defining $\mathfrak{su}(3)$ many-body spin operators to describe the collective spin in ultracold quantum gases composed out of spinful atoms \cite{Heinz2010,Hamley2012,Kawaguchi2012,Stamper2013}, or understanding magnetic impurities in metals by mapping the corresponding local spin operators to so-called Abrikosov fermions \cite{Abrikosov1965}. In quantum optics, it has found applications for defining uncertainty measures \cite{Hertz2019b}, witnessing non-classicality \cite{Arnhem2022,Griffet2023a} or quantum entanglement \cite{Agarwal2005,Griffet2023b} in bosonic systems such as quanta of light \cite{Schleich2001,Mandel2013}.

Despite the Jordan-Schwinger map's importance for relating spin operators with bosonic or fermionic creation and annihilation operators, the relation -- which we refer to as the \textit{basis relation} -- between the induced spin basis and the Fock basis remained incomplete. Simple relations between spin and Fock states are only known for two modes; in this case, the total spin is proportional to the total particle number and its $z$-component to the particle number difference. Finding such relations for arbitrarily many modes is an interesting fundamental problem. This would directly connect discrete with continuous-variable systems, thus allowing one to translate concepts and methods between two intrinsically different kinds of quantum systems. On top of that, such relations are needed when computing the complete measurement statistics of spin observables defined over multiple modes -- as this requires knowing their eigenvectors \cite{Hertz2019b}. 

In this paper, we put forward an algorithmic construction of a complete spin basis for Jordan-Schwinger-type spin operators, which we map to the Fock basis of an arbitrary bosonic or fermionic multi-mode system. Our approach overcomes the problem that working with the total particle number, the total spin, and its $z$-component is, in general, insufficient to faithfully describe all possible Fock states. Although we exclusively consider $\mathfrak{su}(2)$, we stress that our construction extends to other Lie algebras and corresponding quantum systems.

\textit{The remainder of this paper is organized as follows.} In \autoref{sec:ProblemStatement}, we introduce the general Jordan-Schwinger map (\autoref{subsec:JordanSchwingerMap}) together with its archetypal use case, the one-to-one mapping between spin and two-mode bosonic Fock states (\autoref{subsec:FromTwoModesToSpins}), and demonstrate that a straightforward multi-mode extension of the latter fails for both bosons and fermions (\autoref{subsec:BeyondTwoModes}). Then, in \autoref{sec:Construction}, after introducing an extra quantum number -- the counting number $C$ -- in order to lift the degeneracy of the standard spin basis (\autoref{subsec:CountingNumber}), we present our main result -- an algorithmic construction of a complete spin basis over multiple bosonic or fermionic modes (\autoref{subsec:Algorithm}) -- and carry out its initial steps (\autoref{subsec:InitialSteps}). Also, we analyze the degeneracy function and its close link to Gaussian polynomials in the bosonic case (\autoref{subsec:Degeneracy}), as well as non-diagonal spin operators (\autoref{subsec:Unitaries}). We apply our methods to the three-mode case in \autoref{sec:ThreeModes}, leading to explicit basis relations up to $N \le 3$ particles (\autoref{subsec:Setup}) and $N \in \mathbb{N}$ bosons (\autoref{subsec:BasisRelations}), which brings us to a re-interpretation of the genuine tripartite GHZ and W entangled states (\autoref{subsec:TripartiteEntanglement}). Finally, we discuss open questions in \autoref{sec:Conclusion}.

\textit{Notation.} We use natural units $\hbar = 1$ and bold (normal) letters for quantum operators $\boldsymbol{O}$ (classical variables $O$). We sum over indices appearing twice (Einstein's sum convention), with Greek $\alpha, \beta$ and Latin $j,k,l$ indices labeling modes, and basis elements, respectively. The number of modes is denoted as $n$, while $N$ stands for the number of particles -- bosons or fermions.

\section{Problem statement}
\label{sec:ProblemStatement}

\subsection{Jordan-Schwinger map}
\label{subsec:JordanSchwingerMap}
We consider some finite-dimensional Lie algebra $\mathfrak{c}$ with an $n$-dimensional representation $\mathfrak{d}:= \{O_1, ..., O_N \}$ consisting of invertible square matrices $O_j \in \mathfrak{gl}(n,\mathbb{C})$ defined via the Lie bracket
\begin{equation}
    [O_j, O_k] = f_{j k l} \, O_l,
    \label{eq:LieAlgebra}
\end{equation}
where $f_{j k l}$ denote the structure constants. Further, we introduce a set of $n$ independent bosonic or fermionic mode operators endowed with commutation or anticommutation relations
\begin{equation}
    \begin{split}
        [\boldsymbol{a}_{\alpha}, \boldsymbol{a}_{\beta}^{\dagger}] &= \delta_{\alpha \beta} \mathds{1}, \quad [\boldsymbol{a}_{\alpha}, \boldsymbol{a}_{\beta}] = [\boldsymbol{a}_{\alpha}^{\dagger}, \boldsymbol{a}_{\beta}^{\dagger}] = 0, \\
        \{\boldsymbol{a}_{\alpha}, \boldsymbol{a}_{\beta}^{\dagger}\} &= \delta_{\alpha \beta} \mathds{1}, \quad \{\boldsymbol{a}_{\alpha}, \boldsymbol{a}_{\beta}\} = \{\boldsymbol{a}_{\alpha}^{\dagger}, \boldsymbol{a}_{\beta}^{\dagger}\} = 0,
    \end{split}
    \label{eq:ModeOperators}
\end{equation}
respectively, and an associated Fock space $\mathcal{F}$. Then, the algebra $\mathfrak{c}$ can be realized by defining single-body operators according to the generalized Jordan-Schwinger map (also referred to as Schwinger representation)
\begin{equation}
    \begin{split}
        \phi:\mathfrak{gl}(n,\mathbb{C})&\rightarrow \mathfrak{gl}\left( \mathcal{F}\right), \\ O_j &\to \phi (O_j) \equiv \boldsymbol{O}_j = \boldsymbol{a}^{\dagger}_{\alpha} (O_j)_{\alpha \beta} \boldsymbol{a}_{\beta}.
    \end{split}
    \label{eq:JordanSchwingerMap}
\end{equation}
It is straightforward to check that the Jordan-Schwinger map is a homomorphism between representations $\mathfrak{d}$, \textit{i.e.}, that it conserves the Lie bracket \eqref{eq:LieAlgebra}, such that
\begin{equation}
    [O_j, O_k] = f_{j k l} \, O_l \overset{\phi}{\to} [\boldsymbol{O}_j, \boldsymbol{O}_k] = f_{j k l} \, \boldsymbol{O}_l.
\end{equation}
While this work concerns bosonic \textit{or} fermionic systems, we remark that the Jordan-Schwinger map is not restricted to such implementations. More precisely, one may consider any functional combination of bosonic and fermionic operators representing a multimodal algebra. 

\begin{figure*}
    \centering
    \includegraphics[width=0.85\linewidth]{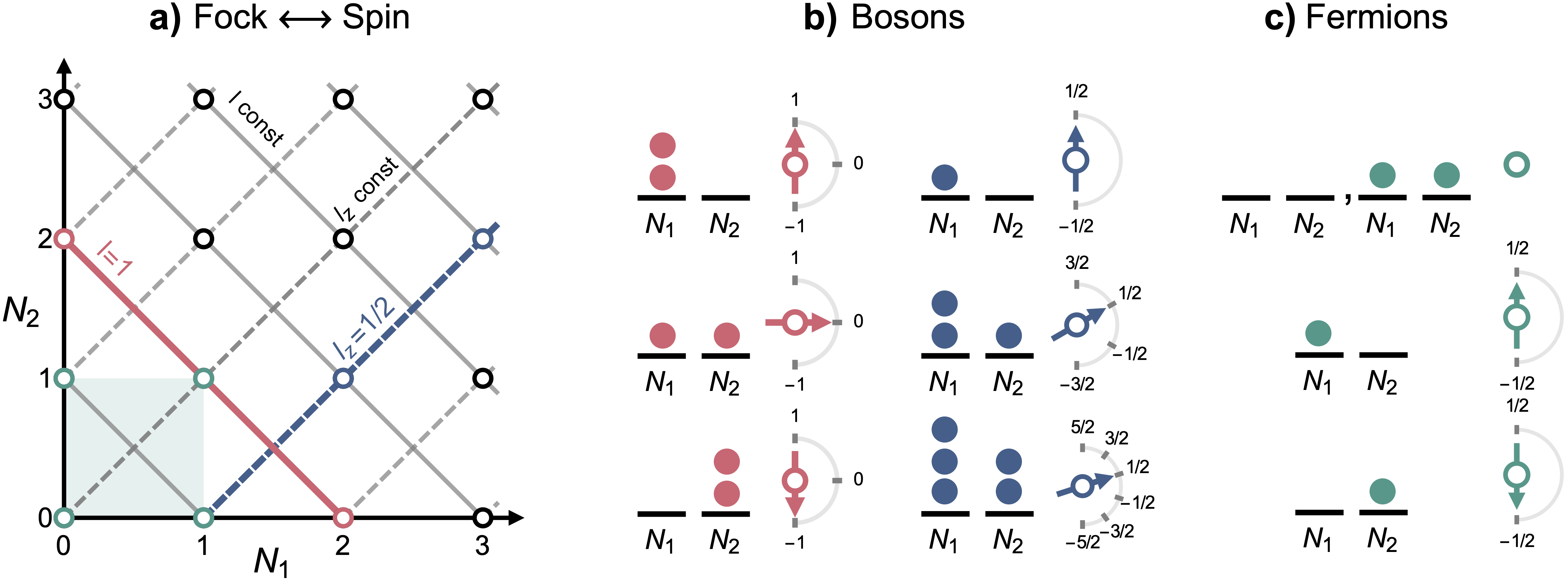}
    \caption{\textbf{a)} Graphical representation of the two-mode Fock basis, with physical states (points) corresponding to integer-valued cartesian coordinates $(N_1,N_2)$, over two bosonic (black and green points) or fermionic (green) modes. For bosons, spin eigenstates follow diagonal (solid) and anti-diagonal (dashed) lines for fixed total spin $l$ and $z$-component $l_z$, respectively. Instead, fermions are constrained by Pauli's principle to $N_1, N_2 = 0,1$ (green square), with fixed $l_z$ still along the anti-diagonal, but $l$ being non-uniquely related to $N_1$ and $N_2$. \textbf{b)} For bosons and $l=1$ (red), the total particle number remains constant $N=2$ when reshuffling particles from the first to the second mode, which leads to the three possible values for $l_z = -1,0,1$. When fixing $l_z = 1/2$ instead (blue), the particle number difference between modes one and two is always one, and adding one particle to each mode increases the total spin by one unit. \textbf{c)} For fermions (green), the two states $(N_1, N_2) = (0,0)$, $(N_1, N_2) = (1,1)$ are both described by $(l,l_z)=(0,0)$ (upper row), while reshuffling particles still corresponds to changing $l_z$ (lower two rows).}
    \label{fig:TwoModes}
\end{figure*}

\subsection{From two modes to spins}
\label{subsec:FromTwoModesToSpins}

The Jordan-Schwinger map is most well-known for its usage in Schwinger's definition of angular momentum in quantum mechanics starting from two (bosonic) modes~\cite{Schwinger1952}. In quantum theory, operators describing the angular momentum or the spin of a particle are defined via representations of the three-dimensional $\mathfrak{su}(2)$ algebra, which is characterized by a totally anti-symmetric structure constant, \textit{i.e.},
\begin{equation}
    [L_j, L_k] = i \, \epsilon_{j k l} \, L_l ,
    \label{eq:su2Algebra}
\end{equation}
where $\epsilon_{j k l}$ denotes the Levi-Civita symbol.

To map this algebra onto two modes, we consider its fundamental representation, which consists of the $2 \times 2$ Pauli matrices
\begin{equation}
    \sigma_x = \begin{pmatrix}
        0 & 1 \\
        1 & 0
    \end{pmatrix}, \,\, \sigma_y = \begin{pmatrix}
        0 & -i \\
        i & 0
    \end{pmatrix}, \,\, \sigma_z = \begin{pmatrix}
        1 & 0 \\
        0 & -1
    \end{pmatrix}, 
\end{equation}
up to a factor one-half, \textit{i.e.}, $L_j = \sigma_j /2$, and define associated spin operators via the Jordan-Schwinger map
\begin{equation}
    \begin{split}
        \boldsymbol{L}_x &\equiv \frac{1}{2} \left(\boldsymbol{a}_1^\dag \boldsymbol{a}_2 + \boldsymbol{a}^\dag_2 \boldsymbol{a}_1 \right),\\
        \boldsymbol{L}_y &\equiv \frac{i}{2} \left(\boldsymbol{a}_2^\dag \boldsymbol{a}_1 - \boldsymbol{a}^\dag_1 \boldsymbol{a}_2 \right),\\
        \boldsymbol{L}_z &\equiv \frac{1}{2} \left(\boldsymbol{a}_1^\dag \boldsymbol{a}_1 - \boldsymbol{a}^\dag_2 \boldsymbol{a}_2 \right).
    \end{split}
    \label{eq:SpinOperatorsTwoModes}
\end{equation}
The corresponding ladder operators read
\begin{equation}
    \boldsymbol{L}_\pm = \frac{1}{\sqrt{2}} \left(\boldsymbol{L}_x \pm i \boldsymbol{L}_y \right),
\end{equation}
which raise or lower the $l_z$ component by one unit since $[\boldsymbol{L}_z, \boldsymbol{L}_\pm] = \pm \boldsymbol{L}_\pm$. We also check that $\boldsymbol{L}_+= \frac{1}{\sqrt{2}} \boldsymbol{a}_1^\dag \boldsymbol{a}_2$ and $\boldsymbol{L}_-= (\boldsymbol{L}_+)^\dag = \frac{1}{\sqrt{2}} \boldsymbol{a}_2^\dag \boldsymbol{a}_1$. The total spin
\begin{equation}
    \boldsymbol{L}^2 = \boldsymbol{L}_x^2 + \boldsymbol{L}_y^2 + \boldsymbol{L}_z^2 = 2 \boldsymbol{L}_+ \boldsymbol{L}_- + \boldsymbol{L}_z \left( \boldsymbol{L}_z - \mathds{1} \right),
\end{equation}
acts as the Casimir operator of $\mathfrak{su}(2)$ and commutes with all generators, that is, $[\boldsymbol{L}^2, \boldsymbol{L}_j] = 0$ for $j \in \{x,y,z \}$. In fact, the total spin and one spin component form a complete set of commuting observables (CSCO). Since $\boldsymbol{L}_z$ is diagonal, we choose the joint eigensystem 
\begin{equation}
    \boldsymbol{L}_z \ket{l, l_z}_s = l_z \ket{l, l_z}_s, \quad \boldsymbol{L}^2 \ket{l, l_z}_s = l (l+1) \ket{l, l_z}_s,
\end{equation}
parameterized by the quantum numbers $l \in \{1/2, 1, ... \}$ and $l_z \in \{-l, -l + 1, ..., l \}$, and we refer to $\{ \ket{l, l_z}_s \}$ as the \textit{spin} basis (note the index $s$).

Choosing the total spin $l$ and its projection $l_z$ amounts to distributing particles over the two modes. Such a mapping requires a relation between the spin basis and the convenient basis for multi-mode systems, which is the \textit{Fock} basis $\{ \ket{N_1, \dots, N_n}_F \}$ spanned by tensor products of single-mode Fock states (note the index $F$)
\begin{equation}
    \ket{N_\alpha}_F = \frac{(\boldsymbol{a}^{\dagger}_{\alpha})^{N_\alpha}}{\sqrt{N_\alpha !}} \ket{0}, 
\end{equation}
with $N_\alpha \in \mathbb{N}$ for bosons and $N_\alpha \in \{ 0, 1\}$ for fermions. Since Fock states are eigenstates of the particle number operator $\boldsymbol{N}_j = \boldsymbol{a}^{\dagger}_j \boldsymbol{a}_j$, that is,
\begin{equation}
    \boldsymbol{N}_{\alpha} \ket{N_{\alpha}}_F = N_{\alpha} \ket{N_{\alpha}}_F,
\end{equation}
the particle numbers $N_1, \dots N_n$ constitute the quantum numbers from the mode picture. 

\subsubsection{Bosons}
When considering two bosons, both $\boldsymbol{L}_z$ and $\boldsymbol{L}^2$ admit a simple rewriting in terms of particle numbers only. Upon defining the total particle number operator as $\boldsymbol{N} = \boldsymbol{N}_1 + \boldsymbol{N}_2$, we find the relations
\begin{equation}
    \boldsymbol{L}_z = \frac{\boldsymbol{N}_1 - \boldsymbol{N}_2}{2}, \quad \boldsymbol{L}^2 = \frac{\boldsymbol{N}}{2} \left(\frac{\boldsymbol{N}}{2} + \mathds{1} \right).
    \label{eq:BosonicOperatorRelationTwoModes}
\end{equation}
They point out a deep connection between the spin and bosonic degrees of freedom: the first equation relates particle number differences with the $\boldsymbol{L}_z$ operator, and the second relation shows that the eigenspaces of total spin operator $\boldsymbol{L}^2$ and total number operator $\boldsymbol{N}$ are equivalent. Hence, the basis relations between Fock and spin states are given by
\begin{equation}
    \begin{split}
        \ket{l,l_z}_s &= \ket{l+l_z,l-l_z}_F, \\
        \ket{N_1,N_2}_F &= \Ket{\frac{N_1+N_2}{2},\frac{N_1-N_2}{2}}_s.
    \end{split}
    \label{eq:BosonicBasisRelationTwoModes}
\end{equation}
These relations can be depicted graphically on a square lattice over the particle numbers $(N_1, N_2)$, see \autoref{fig:TwoModes}~\textbf{a)}. In this representation, the eigenspaces of $\boldsymbol{L}^2$ and $\boldsymbol{L}_z$ correspond to straight lines along the diagonal and anti-diagonal elements, respectively, such that indeed keeping the quantum number $l$ or $l_z$ constant requires a fixed total number of particles or particle number difference, respectively.

\subsubsection{Fermions}
An analogous calculation for fermions reveals that
\begin{equation}
    \boldsymbol{L}_z = \frac{\boldsymbol{N}_1 - \boldsymbol{N}_2}{2}, \quad \boldsymbol{L}^2 = 3\,\frac{\boldsymbol{N}}{2} \left( \mathds{1} -\frac{\boldsymbol{N}}{2}\right).
    \label{eq:FermionicOperatorRelationTwoModes}
\end{equation}
While the $z$-component $\boldsymbol{L}_z$ does still correspond to the particle number difference, the total spin $\boldsymbol{L}^2$ is only \textit{injectively} related to the total particle number $\boldsymbol{N}$, thereby forbidding inversion in general. Strikingly,  two Fock states $\ket{0,0}_F$ and $\ket{1,1}_F$ [see the two diagonal vertices on the green square in \autoref{fig:TwoModes} \textbf{a)} and the upper row in \textbf{c)}] are indistinguishable in the spin basis as they both exhibit vanishing total spin $l=0$ as well as projection $l_z=0$.

We conclude that $\{\boldsymbol{L}^2, \boldsymbol{L}_z\}$ is \textit{not} a CSCO for the Fock space over two fermions. Nevertheless, we can complete it with the operator $\boldsymbol{N}$, which leads to the basis relations
\begin{equation}
    \begin{split}
        \ket{N,l,l_z}_s &= \ket{l+l_z,l-l_z}_F, \\
        \ket{N_1,N_2}_F &= \Ket{N_1+N_2,\frac{N_1\oplus N_2}{2},\frac{N_1-N_2}{2}}_s,
    \end{split}    
    \label{eq:FermionicBasisRelationTwoModes}
\end{equation}
where we used the symbol $\oplus$ for the boolean \texttt{OR} operator. Note that this comes at the price of working with a non-minimal set of commuting operators $\{\boldsymbol{N},\boldsymbol{L}^2, \boldsymbol{L}_z\}$ due to the functional relation between $\boldsymbol{N}$ and $\boldsymbol{L}^2$.

\subsection{Beyond two modes -- a dimensional analysis}
\label{subsec:BeyondTwoModes}
Given the bosonic relations \eqref{eq:BosonicOperatorRelationTwoModes} and \eqref{eq:BosonicBasisRelationTwoModes} as well as their fermionic analogs \eqref{eq:FermionicOperatorRelationTwoModes} and \eqref{eq:FermionicBasisRelationTwoModes}, one might expect similar relations beyond the two-mode case. As seen from the treatise of fermions, a CSCO with spin operators requires at least the inclusion of a third operator, namely the number operator $\boldsymbol{N}$. This raises the question of whether the set $\{ \boldsymbol{N}, \boldsymbol{L}^2, \boldsymbol{L}_z \}$ suffices to faithfully describe an arbitrary configuration of $N$ particles (bosons or fermions) distributed over $n$ modes, \textit{i.e.}, if it constitutes a CSCO.

Let us formalize this problem. For simplicity, we work with the representations in which the matrix $L_z$ is diagonal (and $L_x,L_y$ tridiagonal), namely
\begin{equation}
    \begin{split}
        (L_x)_{\alpha \beta} &= \chi (n, \alpha, \beta) \left(\delta_{\alpha+1,\beta} + \delta_{\alpha, \beta+1} \right), \\
        (L_y)_{\alpha \beta} &= i \chi (n, \alpha, \beta) \left(\delta_{\alpha+1, \beta} - \delta_{\alpha, \beta+1} \right), \\
        (L_z)_{\alpha \beta} &= \left(\frac{n+1}{2} - \alpha \right) \delta_{\alpha, \beta},
    \end{split}
    \label{eq:LzMatrixDiagonal}
\end{equation}
with $\chi (n, \alpha, \beta) = \left(\sqrt{\frac{n+1}{2}\left(\alpha+\beta-1\right)-\alpha\beta} \,\right)/2$ and $\alpha, \beta \in \{ 1, \dots, n \}$. (Other possible representations are discussed in \autoref{subsec:Unitaries}.) The corresponding quantum operators in the mode picture follow from \eqref{eq:JordanSchwingerMap} and read
\begin{equation}
    \begin{split}
        \boldsymbol{L}_x &= \frac{1}{2} \sum_{j=1}^{n-1}\sqrt{j(n-j)} \left( \boldsymbol{a}_{j+1}^\dagger \boldsymbol{a}_j + \boldsymbol{a}_j^\dagger \boldsymbol{a}_{j+1}  \right), \\
        \boldsymbol{L}_y &=  \frac{i}{2} \sum_{j=1}^{n-1}\sqrt{j(n-j)} \left( \boldsymbol{a}_{j+1}^\dagger \boldsymbol{a}_j - \boldsymbol{a}_j^\dagger \boldsymbol{a}_{j+1} \right), \\
        \boldsymbol{L}_z &= \sum_{j=1}^n \left( \frac{n+1}{2}-j \right) \boldsymbol{N}_j,
    \end{split}
    \label{eq:LzOperatorDiagonal}
\end{equation}
with ladder operators
\begin{equation}
    \begin{split}
        \boldsymbol{L}_+ &= \frac{1}{\sqrt{2}}\sum_{j=1}^{n-1}\sqrt{j (n-j)} \, \boldsymbol{a}_{j+1}^\dagger \boldsymbol{a}_j, \\
        \boldsymbol{L}_- &= \frac{1}{\sqrt{2}}\sum_{j=1}^{n-1}\sqrt{j (n-j)} \, \boldsymbol{a}_j^\dagger \boldsymbol{a}_{j+1},
    \end{split}
\end{equation}
and total spin
\begin{equation}
    \begin{split}
        \boldsymbol{L}^2 &= \left[\sum_{j=1}^n\left(\frac{n+1}{2}-j\right)\boldsymbol{N}_j\right]^2\\
        &\hspace{0.4cm}+\sum_{j=1}^n \left[(n-1) \left(j-\frac{1}{2} \right) - j^2 \right] \boldsymbol{N}_j\\
        &\hspace{0.4cm}\pm \sum_{j=1}^{n-1}\sum_{k=1}^{n-1}\sqrt{j(n-j)k(n-k)} \, \boldsymbol{a}_{j+1}^\dagger\boldsymbol{a}_{k}^\dagger\boldsymbol{a}_j\boldsymbol{a}_{k+1},
    \end{split}
\end{equation}
with $+$ and $-$ for bosons and fermions, respectively.

On one hand, in the mode picture, the basis states describing configurations of $N$ particles over $n$ modes correspond to the eigenspace of $\boldsymbol{N}$, which we denote as $\mathcal{E}_N = \{ \ket{\psi}: \boldsymbol{N} \ket{\psi} = N \ket{\psi} \}$. Note here that $\mathcal{E}_N$ is a submodule of the $\mathfrak{su}(2)$ algebra since $\boldsymbol{N}$ is the image of the Jordan-Schwinger map applied to the identity $\mathds{1}$ and thus indeed commutes with all operators generated by the Jordan-Schwinger map -- elements of $\mathfrak{su}(2)$ included. On the other hand, in the spin picture, we have the set of basis states described by the tuple $(N, l, l_z)$ for fixed $n$, which we denote by $\Gamma_N = \{ (N,l,l_z) \}$, where
\begin{equation}
    \begin{split}
        l_z &= \sum_{j = 1}^n \left( \frac{n+1}{2} - j \right) N_j \\
        &= \frac{n-1}{2} \, N_1 + \frac{n-3}{2} \, N_2 + \cdots +  \frac{1-n}{2} \, N_n.
    \end{split}
    \label{eq:lzDiagonal}
\end{equation}
Thus, we can reformulate the above problem as a combinatorial problem: Given a fixed number of particles $N$ over $n$ modes, can we always ensure that $ \# \Gamma_N \ge \dim \mathcal{E}_N$? Since particle nature influences counting possible states, we distinguish bosons and fermions in the following.

\subsubsection{Bosons}
All $N$ particles can reside in any of the $n$ modes for bosons. Taking their indistinguishability into account, the number of basis states of $\mathcal{E}_N$ is
\begin{equation}
    \dim \mathcal{E}_N = {n+N-1 \choose N } \sim \mathcal{O}(N^{n-1}).
    \label{eq:dimWBosons}
\end{equation}
The dimension of $\Gamma_N$ follows from basic algebraic properties of $\mathfrak{su}(2)$. First, we recall that $0 \le l \le \max l_z$ with $2l+1$ eigenvalues for $l_z$ given some $l$. Then, for $N$ bosons, the quantum number $l_z$ is maximized when all $N$ bosons reside in the first mode as is clear from Eq. \eqref{eq:lzDiagonal}, see \autoref{fig:TwoModes} \textbf{b)}. This results in a maximum value of $N(n-1)/2$ for $l$, and hence the dimension of $\Gamma_N$ is given by
\begin{equation}
    \begin{split}
        \hspace{-0.05cm} \# \Gamma_N &= \sum_{l=0,\frac{1}{2},\dots}^{N(n-1)/2} (2 l+1) \\
        &= \frac{\left[1 + (n-1) N \right] \left[2 + (n-1) N \right]}{2} \sim \mathcal{O}(N^2).
    \end{split}
    \label{eq:dimGammaBosons}
\end{equation}
Interestingly, \eqref{eq:dimWBosons} and \eqref{eq:dimGammaBosons} are monotonic in both $n$ and $N$. We discover the inclusion of the third operator $\boldsymbol{N}$ to be sufficient for the spin basis to encompass up to the three-mode bosonic Fock basis since $\# \Gamma_N \ge \dim \mathcal{E}_N$ for all $N$ when $n \le 3$. However, when $n > 3$, it is straightforward to find sufficiently large values for $N$ for which $\# \Gamma_N < \dim \mathcal{E}_N$. For instance, consider $n=4$, in which case $\# \Gamma_N \sim \mathcal{O} (N^2)$ whereas $\dim \mathcal{E}_N \sim \mathcal{O}(N^3)$, with the crossover being located at $N=22$.

\subsubsection{Fermions}
We lay out a similar argument for fermions. In this case, Pauli's principle reduces the dimension of $\mathcal{E}_N$ to
\begin{equation}
    \dim \mathcal{E}_N = {n \choose N },
    \label{eq:dimWFermions}
\end{equation}
with $N \le n$ understood. Similarly, $l_z$ is constrained by $N_j \le 1$, such that its maximum is attained when the first $N$ modes are filled with one particle each, see \eqref{eq:lzDiagonal} and \autoref{fig:TwoModes} \textbf{c)}, in which case $l_z = N (n-N)/2$. Note that $-N$ replaces $-1$ in the corresponding bosonic expression for the maximum $l_z=N(n-1)/2$ (this will be exploited later on to simplify the notation). Thus,
\begin{equation}
    \begin{split}
        \# \Gamma_N &= \sum_{l=0,\frac{1}{2},\dots}^{N(n-N)/2} (2l+1) \\
        &=  \frac{\left[1 + (n-N) N \right] \left[2 + (n-N) N \right]}{2}.
    \end{split}
    \label{eq:dimGammaFermions}
\end{equation}
In contrast to the bosonic case, both functions \eqref{eq:dimWFermions} and \eqref{eq:dimGammaFermions} are symmetric with respect to $N=n/2$ for even $n, N$ when considered as functions of $N$, at which they become maximal. For this value, we find the scalings $\mathcal{E}_{n/2} = \mathcal{O}(2^n/\sqrt{n})$ and $\# \Gamma_{n/2} = \mathcal{O} (n^4)$ when $n$ grows large, such that $\# \Gamma_{N} < \dim \mathcal{E}_N$ for $N=n/2$ and sufficiently large $n$. Note, however, that the turning point is now at $n \ge 12$, thereby allowing a faithful description of Fock states in terms of spin states up to twelve modes. This larger range witnesses the fact that the fermionic configuration space $\mathcal{E}_N$ is much more constrained than its bosonic counterpart given Pauli's principle. Hence, its dimension is much smaller and one needs a larger number of modes $n$ until it exceeds the dimension of $\Gamma_N$. 

We conclude that the set of commuting observables $\{ \boldsymbol{N}, \boldsymbol{L}^2, \boldsymbol{L}_z \}$ is, in general, \textit{insufficient} to single out every Fock state of multi-mode bosonic or fermionic systems. In this sense, the usually considered case of two modes, for which the links between the Fock and pin bases are well-known, constitutes an exception rather than the norm. This raises the quest for a general basis relation between the mode picture, \textit{i.e.}, continuous variables, and the spin picture, \textit{i.e.}, discrete variables. In the next section, we construct such a relation by minimally extending the set of commuting observables $\{ \boldsymbol{N}, \boldsymbol{L}^2, \boldsymbol{L}_z \}$ and including an extra one, whose associated quantum number we call the counting number $C$.

\section{Multi-mode spin basis}
\label{sec:Construction}

\subsection{Counting number $C$}
\label{subsec:CountingNumber}
To represent an arbitrary $n$-mode (bosonic or fermionic) Fock state in terms of a spin basis which includes at least the three quantum numbers $\{N,l,l_z\}$, we introduce one additional quantum number $C$ to discriminate between all states sharing the same three first quantum numbers. This quantum number $C$ is linked to the degeneracy of the value $l_z$ for given $n, N$, which we denote by $\mathfrak{g}(n,N,l_z)$. Formally, we define the latter as the dimension of the intersection of $\mathcal{E}_N$ with the eigenspace $\mathcal{E}_{l_z}$ of $\boldsymbol{L}_z$ corresponding to the eigenvalue $l_z$, namely
\begin{equation}
    \mathfrak{g}(n,N,l_z) = \dim \mathcal{E}_{N, l_z}.
\end{equation}
We again find an explicit expression using combinatorics: $\mathfrak{g}(n,N,l_z)$ corresponds to the number of configurations of $N$ particles in $n$ modes with a ``score'' $l_z$ as given by Eq.~\eqref{eq:lzDiagonal}, that is,
\begin{equation}
    \begin{split}
        \mathfrak{g} (n,N,l_z) &= \# \Bigg\{ N_j \in \mathbb{N}: \sum_{j=1}^{n} N_j = N, \\ 
        &\hspace{1.1cm}\sum_{j=1}^{n} \left( \frac{n+1}{2} - j \right) N_j = l_z \Bigg\}.
    \end{split}
    \label{eq:DegeneracyDefinition}
\end{equation}
While this suffices to describe degeneracy in the bosonic case, fermions are additionally constrained by $N_j \le 1$. We discuss the degeneracy function $\mathfrak{g}(n,N,l_z)$ and its link to  Gaussian polynomials in detail in \autoref{subsec:Degeneracy}.

\subsection{General algorithm}
\label{subsec:Algorithm}
We are now ready to state our main result, namely an algorithm for generating an extended spin basis that admits a basis relation with the Fock basis in the mode picture. For a fixed number of modes $n$ and particles $N$, we construct such a complete basis $B$ using the four quantum numbers $\{N,l,l_z, C \}$. The basic idea is to identify the kernel of the raising operator $\ker \boldsymbol{L}_+$ on $\mathcal{E}_{N,l}$, which gives the highest-weight state $\ket{N,l,l}_s$ in terms of Fock states. Smaller values for the $z$-component $l_z < l$ then follow by repeatedly applying the lowering operator $\boldsymbol{L}_-^{l-l_z}$. Algorithmically, we implement this procedure as follows.
\begin{enumerate}
    \item \textbf{Maximum $l_z$}. We construct the first basis element by identifying the unique highest weight state with maximum $l_z = l = N (n-\kappa)/2$ with $\kappa = 1$ or $\kappa = N$ for bosons or fermions, respectively. For bosons, this state corresponds to all particles residing in the first mode $\ket{N,0,\dots,0}_F$, while for fermions, the first $N$ modes are filled with one particle each $\ket{1,\dots,1,0, \dots, 0}_F$. In the spin basis, it reads
    \begin{equation}
        B_{l} = \left\{ \ket{N,N\frac{n-\kappa}{2},N\frac{n-\kappa}{2},0}_s \right\},
    \end{equation}
where the counting number $C$ is arbitrarily set to zero for the first basis element.
    
    \item \textbf{Non-negative $l_z$}. Then, we devise a basis for every fixed value of $l_z$ beginning with $l_z = l - 1$ up to $l_z = 0$ or $1/2$ for an even or odd total spin, respectively, in unit steps:
    \begin{enumerate}
        \item We consider the $\mathfrak{g}(n,N,l_z)$ vectors $\ket{\psi}$ with eigenvalue $l_z$. These span $\mathcal{E}_{N,l_z}$.
        \item We obtain the first part of the corresponding basis by applying the lowering operator on all basis vectors corresponding to the basis of $\left(l_z+1\right)$, \textit{i.e.},
        \begin{equation}
            \hspace{1cm}B_{l_z}^0=\left\{\frac{\boldsymbol{L}_-\ket{\psi}}{\left\|\boldsymbol{L}_-\ket{\psi}\right\|}, \, \forall\ket{\psi}\in B_{l_z+1}\right\}.
        \end{equation}
        This set spans $\boldsymbol{L}_-\mathcal{E}_{N,l_z+1}$ and contains $\mathfrak{g}(n,N,l_z+1)$ vectors.
        \item We construct $\mathfrak{g}(n,N,l_z)-\mathfrak{g}(n,N,l_z+1)$ vectors orthonormal to each other and to $B_{l_z}^0$ with $l = l_z$ from $\mathcal{E}_{N,l_z}$ (for instance, using the Gram-Schmidt algorithm), which form the set
        \begin{equation}
            B_{l_z}^1=\left\{\ket{N,l_z,l_z,C}_s \right\},
        \end{equation}
        where the counting number sweeps the range $C \in \{0,1,\dots, \mathfrak{g}(n,N,l_z)-\mathfrak{g}(n,N,l_z+1) - 1\}$, and span $\mathcal{E}_{N,l_z}\setminus \boldsymbol{L}_-\mathcal{E}_{N,l_z+1}$.
        \item The full basis of $\mathcal{E}_{N,l_z}$ follows by addition
        \begin{equation}
            B_{l_z}=B_{l_z}^0+B_{l_z}^1.
        \end{equation}
    \end{enumerate}
    \item \textbf{Negative $l_z$}. For negative $l_z$, we apply the lowering operator to all non-vanishing vectors corresponding to $l_z+1$
    \begin{widetext}
        \begin{equation}
            B_{l_z}=\left\{\frac{\boldsymbol{L}_-\ket{N,l,l_z+1,C}_s}{\left\|\boldsymbol{L}_-\ket{N,l,l_z+1,C}_s\right\|},\forall\ket{N,l,l_z+1,C}_s\in B_{l_z+1}, \abs{l_z+1}\neq l\right\}.
        \end{equation}
    \end{widetext}
\end{enumerate}

\begin{figure*}
    \centering
    \includegraphics[width=0.95\linewidth]{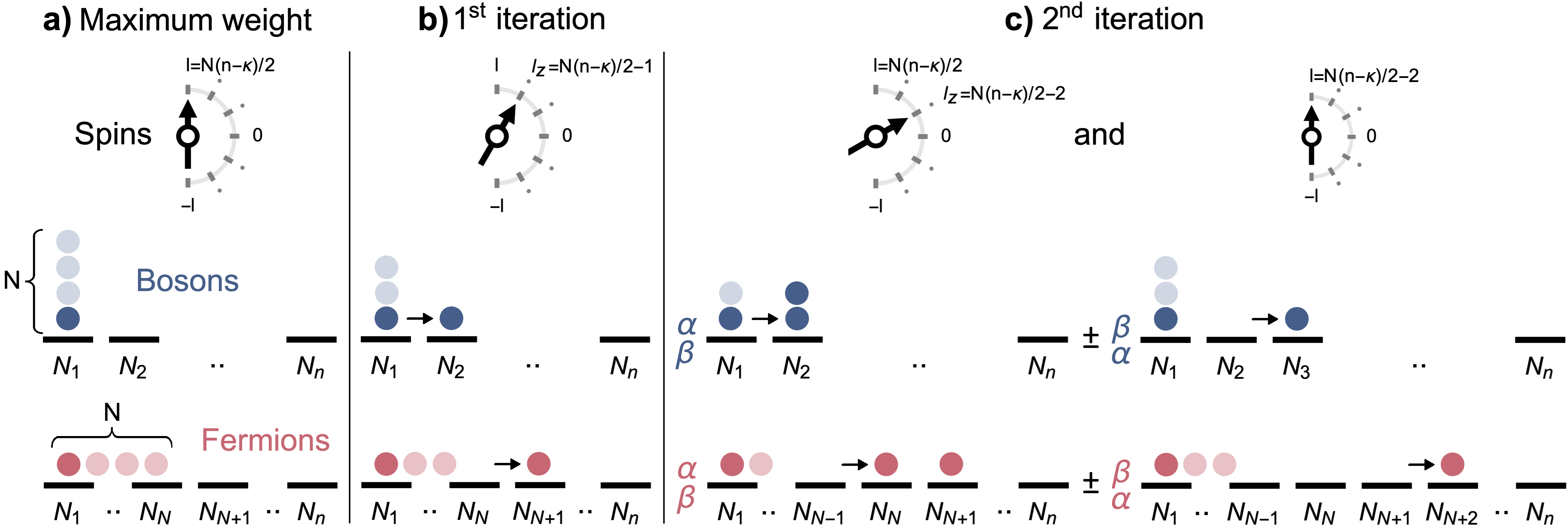}
    \caption{Construction of the spin bases as described in \autoref{subsec:InitialSteps}. \textbf{a)} For given numbers of modes $n$ and particles $N$, the maximum weight state of the spin basis is attained for $l_z = l = N (n-\kappa)/2$, with $\kappa = 1$ for bosons and $\kappa = N$ for fermions, respectively. This state is represented by either $N$ bosons (blue) residing in the first mode or the first $N$ modes being populated by one fermion (red) each. \textbf{b)} The next (unique) spin state follows by applying the lowering operator $\boldsymbol{L}_-$, which results in $l_z = N(n-\kappa)/2 - 1$ and $l=N (n-\kappa)/2$ left unchanged. This corresponds to one of the first mode's bosons or the $N$th fermion being shifted to the second or the $N+1$th mode, respectively. \textbf{c)} Applying $\boldsymbol{L}_-$ once more gives the state corresponding to $l_z = N(n-\kappa)/2 - 2$ (left), to which a second spin state with $l_z = l = N (n-\kappa)/2-2$ (right) is orthogonal. Both are linearly related to the two Fock states generated by the application of $\boldsymbol{L}_-$, which correspond to two bosons in the second mode (left) or one boson in the third mode (right), as well as one fermion each in the $N$th and the $(N+1)$th mode (left) and one fermion in the $(N+2)$th mode (right), respectively.}
    \label{fig:Algorithm}
\end{figure*}

\subsection{Initial steps}
\label{subsec:InitialSteps}
To illustrate the algorithm, we outline its first steps for an arbitrary number of modes $n$ and particles $N$. We start from the highest weight state with $l_z = l = N(n-\kappa)/2$, see \autoref{fig:Algorithm} \textbf{a)}, which fulfills
\begin{equation}
    \begin{split}
        &\ket{N,N\frac{n-\kappa}{2},N\frac{n-\kappa}{2},0}_s \\
        &= \begin{cases}
        \ket{N, 0, \dots}_F &\text{bosons}, \\
        \ket{1, \dots, 1, 0, \dots}_F &\text{fermions},
    \end{cases}
    \end{split}
    \label{eq:FirstStepsl}
\end{equation}
with $\kappa = 1$ for bosons (blue) and $\kappa = N$ for fermions (red), respectively. Since the quantum number $l_z$ is always decreased by one unit in the following, the maximum weight state also specifies whether $l_z$ runs in integer or half-integer values. Interestingly, $l_z \in \mathbb{Z}$ is guaranteed if $N$ is even (see \autoref{fig:TwoModes}, left panel), but also holds if both $n$ and $N$ are odd. In contrast, half-integer values $l_z \in \mathbb{Z} + 1/2$ are attained if $n$ is even and $N$ is odd [see \autoref{fig:TwoModes}, right panel in \textbf{b)} and two lower rows in \textbf{c)}].

Now, we lower the $z$-component by one unit to $l_z=N(n-\kappa)/2-1$ by applying the lowering operator, see \autoref{fig:Algorithm} \textbf{b)}, which results in
\begin{equation}
    \begin{split}
        &\ket{N,N\frac{n-\kappa}{2},N\frac{n-\kappa}{2}-1,0}_s \\
        &\equiv\frac{\boldsymbol{L}_-\ket{N,N\frac{n-\kappa}{2},N \frac{n-\kappa}{2},0}_s}{\left\|\boldsymbol{L}_-\ket{N,N\frac{n-\kappa}{2},N\frac{n-\kappa}{2},0}_s\right\|}.
    \end{split}
    \label{eq:FirstStepslm1a}
\end{equation}
As the dimension of the corresponding subspace is $\mathfrak{g}(n,N,N(n-\kappa)/2-1)=1$, there is precisely one configuration for both particle types realizing this value for $l_z$. For bosons, one excitation in the first mode is transferred to the second mode. In contrast, for fermions, the excitation in the $N$th mode is shifted to the $(N+1)$th mode\footnote{The required normalization, in this case, is $\sqrt{\kappa (n-1})$, which is implied by how $\boldsymbol{L}_-$ acts onto the highest weight state.}. Therefore, we can make the assignment
\begin{equation}
    \begin{split}
        &\ket{N,N\frac{n-\kappa}{2},N\frac{n-\kappa}{2}-1,0}_s \\
        &= \begin{cases}
            \ket{N-1, 1, 0, \dots}_F & \text{bosons}, \\
            \ket{1, \dots, 1, 0, 1, 0, \dots}_F & \text{fermions}.
        \end{cases}
    \end{split}
    \label{eq:FirstStepslm1b}
\end{equation}
One may verify this directly by applying the Jordan-Schwinger expression of the lowering operator $\boldsymbol{L}_-$ to the Fock states in Eq. \eqref{eq:FirstStepsl}.

Next, we consider the value $l_z=N(n-\kappa)/2-2$, for which we need to construct $\mathfrak{g}(n,N,N(n-\kappa)/2-2)=2$ vectors\footnote{For bosons, one vector is sufficient if $N=1$ or $n=2$, while zero vectors remain when both conditions are fulfilled. One vector is needed for fermions when $N=1$ or $N=n-1$ for $n \ge 3$.}. As before, we first apply the lowering operator to the only vector with $l_z=N(n-\kappa)/2-1$ given by \eqref{eq:FirstStepslm1a}, leading to
\begin{equation}
    \begin{split}
        &\ket{N,N\frac{n-\kappa}{2},N\frac{n-\kappa}{2}-2,0}_s \\
        &\equiv\frac{\boldsymbol{L}_-\ket{N,N\frac{n-\kappa}{2},N\frac{n-\kappa}{2}-1,0}_s}{\left\|\boldsymbol{L}_-\ket{N,N\frac{n-\kappa}{2},N\frac{n-\kappa}{2}-1,0}_s\right\|},
    \end{split}
    \label{eq:FirstStepslm2a}
\end{equation}
see left spin in \autoref{fig:Algorithm} \textbf{c)}. In simple words, the lowering operator $\boldsymbol{L}_-$ acts on Fock states by shuffling particles from \textit{every} $j$th to the $(j+1)$th mode. Thus, the latter state reduces to a linear combination of the Fock states $\ket{N-2,2,0 \dots}_F$ and $\ket{N-1,0,1,0,\dots}_F$ in the bosonic case, and $\ket{1, \dots, 0, 1, 1, 0, \dots}_F$ and $\ket{1, \dots, 0, 0, 0, 1, 0, \dots}_F$ in the fermionic case. In both cases, the precise coefficients $\alpha, \beta$ depend on the mode and particle numbers. Note that this construction continues for smaller values of $l_z$, thereby enlarging the set of corresponding Fock states in each step, which is in contrast to the first two vectors of the spin basis representing each a unique Fock state, see \eqref{eq:FirstStepsl} and \eqref{eq:FirstStepslm1b}. The second vector associated with two quanta of excitation is constructed by setting $l=l_z=N(n-\kappa)/2-2$ in the latter expression, namely
\begin{equation}
    \ket{N,N\frac{n-\kappa}{2}-2,N\frac{n-\kappa}{2}-2,0}_s,
    \label{eq:FirstStepslm2b}
\end{equation}
see right spin in \autoref{fig:Algorithm} \textbf{c)}. It is orthonormal to the first vector \eqref{eq:FirstStepslm2a} by definition as its total spin quantum number differs and can be linearly decomposed into the two Fock states above by interchanging the prefactors $\alpha, \beta$ and reversing one sign.

\subsection{Degeneracy function}
\label{subsec:Degeneracy}

\subsubsection{Recursion relations}
We found the degeneracy function $\mathfrak{g}(n,N,l_z)$ to play a crucial role in constructing the extended spin bases following our algorithm. Starting from its implicit definition \eqref{eq:DegeneracyDefinition}, we will now derive recurrence relations to assess the degeneracy by simple means. For clarity, we first introduce the function
\begin{equation}
    \begin{split}
        \mathfrak{h} (n,N,h) &= \# \Bigg\{ N_j \in \mathbb{N}: \sum_{j=1}^{n} N_j = N, \sum_{j=1}^{n} j N_j = h \Bigg\},
    \end{split}
    \label{eq:hDefinition}
\end{equation}
such that $\mathfrak{g}(n,N,l_z) = \mathfrak{h} (n,N,N(n+1)/2 - l_z)$. We refer to the value of $h$ as the ``score'' of a configuration $\{N_j\}$. Then, we divide the number of configurations counted by $\mathfrak{h}$ into two subsets: those with no particles in the first mode and those with at least (precisely) one bosonic (fermionic) particle in the first mode, respectively. We thus effectively distribute the $N$ particles over one mode less $n-1$ in the former case (when the first mode is empty). After relabeling the remaining modes, we find this to be equivalent to reducing the score by $N$, such that there are $\mathfrak{h} (n-1,N,h-N)$ configurations left, independent of the particles' nature. 

In the latter case (when the first mode is not empty), a distinction between bosons and fermions becomes necessary. For bosons, the corresponding number of configurations is $\mathfrak{h}(n,N-1,h-1)$, as at least one particle is guaranteed to reside in the first mode so that there remain $N-1$ particles to distribute over the $n$ modes, with this particle's contribution reducing the score to $h-1$. For fermions, however, Pauli's principle forbids more than one fermion in the first mode, such that the remaining $N-1$ particles have to be distributed over the residual $n-1$ modes. As before, excluding one mode reduces the score by $N$ after relabeling the modes, resulting in $\mathfrak{h}(n-1, N-1,h-N)$ configurations for fermions.

\begin{figure*}
    \centering
    \includegraphics[width=0.99\linewidth]{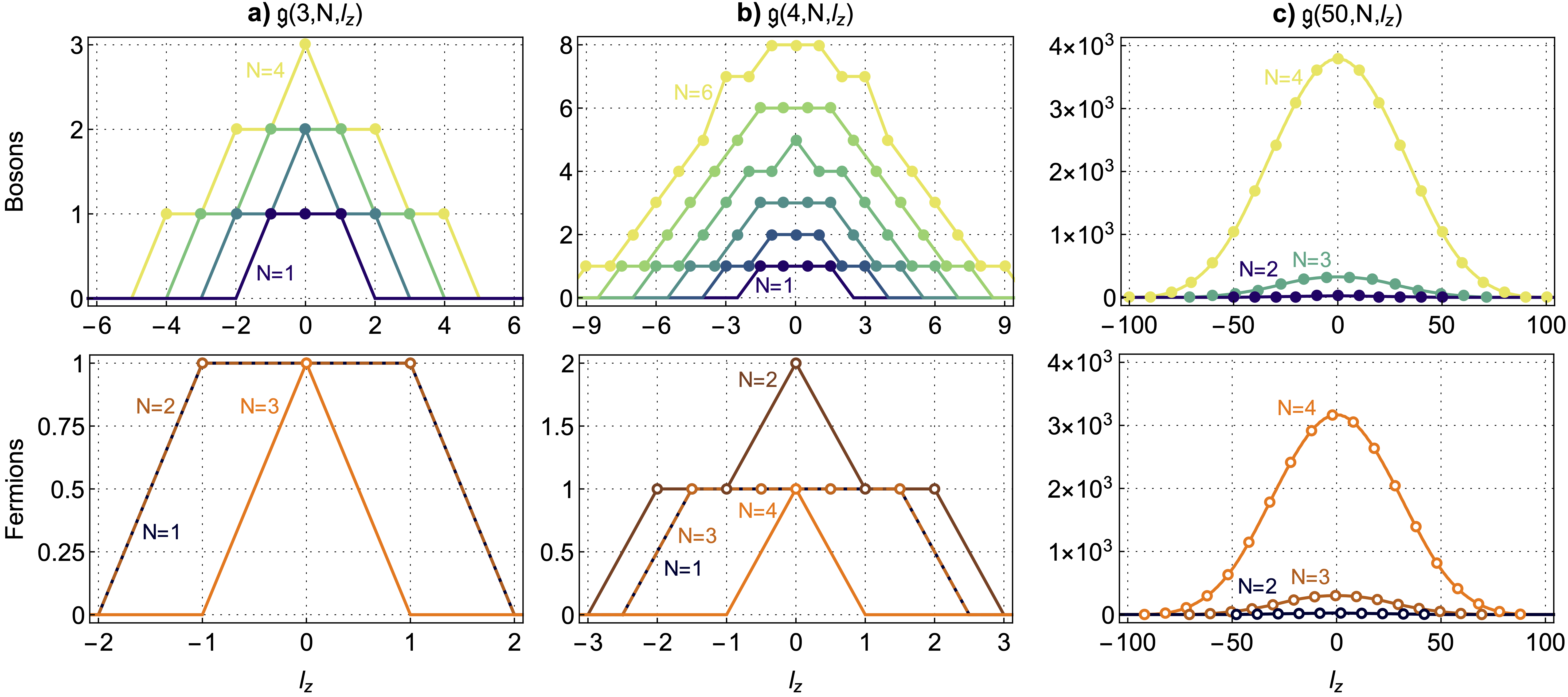}
    \caption{Bosonic (upper row) and fermionic (lower row) degeneracy $\mathfrak{g}(n,N,l_z)$ as function of $l_z$ for $n=3$, $n=4$ and $n=50$ modes in \textbf{a)}, \textbf{b)} and \textbf{c)}, respectively, for various particle numbers $N$.}
    \label{fig:Degeneracy}
\end{figure*}

Since the division into the two subsets is complete, we end up with the two recursion formulas
\begin{equation}
    \begin{split}
        \mathfrak{h}(n,N,h) &= \mathfrak{h} (n-1,N,h-N) \\
        &\hspace{0.25cm}+ \begin{cases}
            \mathfrak{h} (n,N-1,h-1) & \text{bosons}, \\
            \mathfrak{h} (n-1,N-1,h-N) & \text{fermions}.
        \end{cases}
    \end{split}
    \label{eq:RecursionFormulas}
\end{equation}

It is left to specify initial conditions. If there are no particles to distribute ($N=0$), the score has to vanish ($h=0$), which holds for both particle types. Additionally, for a single bosonic mode ($n=1$), all $N$ particles have to reside in this mode, necessarily giving a score of $N$. Thus, the bosonic initial conditions read
\begin{equation}
    \begin{split}
        \delta_{h , 0} = \mathfrak{h}(n,0,h) = \mathfrak{h} (1,N,h+N).
    \end{split}
    \label{eq:BosonicInitialConditions}
\end{equation}

For fermions, we must respect Pauli's principle: the admissible score $h$ is bounded from below by $N(N+1)/2$, which is realized when the first $N$ modes are filled. Analogously, $h$ has the upper bound $N(1-N)/2+nN$, for which the last $N$ modes are populated, such that\footnote{Fermions are further restricted by the trivial condition $N \le n$, whose implementation can improve the recursion relation's computing time.}
\begin{equation}
    \begin{split}
        1 &= \mathfrak{h} (n,N,N (N+1)/2) \\
        &= \mathfrak{h} (n,N,N(1-N)/2+nN), \\
        0 &= \mathfrak{h} (n,N,h) \text{ if } h < N (N+1)/2 \\
        &\hspace{1.86cm}\text{ or } h > N(1-N)/2+nN.
    \end{split}
    \label{eq:FermionicInitialConditions}
\end{equation}

We show the resulting degeneracy functions in \autoref{fig:Degeneracy}. The minimal non-trivial extension of the Jordan-Schwinger map comprises $n=3$ modes, see \autoref{fig:Degeneracy} \textbf{a)}, in which case $l_z$ is always even. Since two consecutive degeneracy values differ by one at most, the quantum number $C$ is always zero, thereby confirming the earlier finding that the set $\{\boldsymbol{N}, \boldsymbol{L}, \boldsymbol{L}_z \}$ is indeed a CSCO in this case. Further, the bosonic degeneracy always grows monotonically with the particle number $N$. In contrast, fermionic degeneracy is maximal around $N \approx n/2$, thereby allowing the degeneracies of small and large particle numbers to coincide, see, \textit{e.g.}, the curves corresponding to $N=1$ (straight, black) and $N=2$ (dashed, dark brown). 

We depict the four-mode case $n=4$ in \autoref{fig:Degeneracy} \textbf{b)}, with the main difference to $n=3$ modes being that the $z$-component $l_z$ attains odd values for odd particle numbers. Also, the bosonic degeneracy jumps by two for $l_z = \pm 4$ when $N = 6$ (see straight yellow curve), which demonstrates the insufficiency of the set of three operators $\{\boldsymbol{N}, \boldsymbol{L}, \boldsymbol{L}_z \}$ for faithfully describing arbitrary Fock states beyond three bosonic modes. Note that the fermionic counterpart does not display this behavior. 

For large mode numbers, see \autoref{fig:Degeneracy} \textbf{c)} for $n=50$, the degeneracy functions, when normalized, do \textit{not} tend to binomial or Gaussian distributions: the corresponding Gaussian curves with the same variances are slightly more peaked, as implied by the maximum entropy principle.

\subsubsection{Link to Gaussian polynomials}
While the bosonic recursion relation \eqref{eq:RecursionFormulas} together with the corresponding initial conditions \eqref{eq:BosonicInitialConditions} suffice to compute the bosonic degeneracy function, they do not provide a deeper understanding, especially in light of the seemingly irregular behavior observed in \autoref{fig:Degeneracy}. Remarkably, one can establish a relation between $\mathfrak{g}(n,N,l_z)$ and Gaussian polynomial coefficients, which represent the $q$-analogs of the well-known binomial coefficients (the usual binomial coefficients are obtained back in the limit $q \to 1$). They read \cite{Majid2002}
\begin{equation}
    \begin{bmatrix}
        j \\ k
    \end{bmatrix}_q = \frac{[j]_q!}{[k]_q! \, [n-k]_q!},
    \label{eq:GaussianPolynomials}
\end{equation}
with natural numbers $k \le j$ and the $q$-factorial and $q$-numbers being defined as
\begin{equation}
    [j]_q ! = \prod_{i=1}^j [i]_q, \quad [j]_q = \sum_{i=0}^{j-1} q^i,
\end{equation}
respectively. Equivalently, Gaussian polynomials are defined via the recursion relation
\begin{equation}
    \begin{bmatrix}
        j \\ k
    \end{bmatrix}_q =
    \begin{bmatrix}
        j-1 \\ k-1
    \end{bmatrix}_q + q^k
    \begin{bmatrix}
        j-1 \\ k
    \end{bmatrix}_q,
    \label{eq:GaussianPolynomialsRecursionRelation}
\end{equation}
with the initial conditions
\begin{equation}
    \begin{bmatrix}
        j \\ j
    \end{bmatrix}_q = 
    \begin{bmatrix}
        j \\ 0
    \end{bmatrix}_q = 1,
    \label{eq:GaussianPolynomialsInitialConditions}
\end{equation}
which is nothing but the $q$-analog of Pascal's identity.

We find that Gaussian polynomials admit a series representation in $q$, whose coefficients can be expressed in terms of the degeneracy function $\mathfrak{h} (n,N,h)$, namely
\begin{equation}
     \begin{bmatrix}
        j \\ k
    \end{bmatrix}_q = \sum_{i=0}^{k (j-k)} \mathfrak{h} (j-k+1, k, i+k) \, q^i.
    \label{eq:GaussianPolynomialsDegeneracyRelation}
\end{equation}
We prove this fundamental relation in \hyperref[app:GaussianPolynomials]{Appendix A}. The close link between Gaussian polynomials and the degeneracy function is rooted in the critical role played by the Gaussian polynomials in solving a variant of the balls into bins problem. The coefficient of $q^N$ in $[j + n \, n]_q$ counts the number of configurations of putting $N$ balls into $n$ indistinguishable bins, with the constraint that each bin carries at most $j$ balls. According to the above relation, this number equals $\mathfrak{h} (j+1, n, N+n)$, and hence is equivalent to counting the number of ways one can distribute $n$ indistinguishable particles over $j+1$ modes with a total score of $h=n+N$.

Unfortunately, we have not found a fermionic counterpart to Eq. \eqref{eq:GaussianPolynomialsDegeneracyRelation}, where the fermionic degeneracy functions that are solutions of the recursion relations \eqref{eq:RecursionFormulas} with initial conditions \eqref{eq:FermionicInitialConditions} would possibly appear in another polynomial series.

\subsection{Other spin operators}
\label{subsec:Unitaries}
So far, we have derived relations between the Fock and spin bases for the case that the matrix $L_z$ is diagonal. We now generalize our construction to other spin observables such as, \textit{e.g.}, $\boldsymbol{L}_x$ or $\boldsymbol{L}_y$, and also other representations of $\mathfrak{su}(2)$. To this end, we consider any hermitian Jordan-Schwinger operator $\boldsymbol{L}_j = \phi(L_j)$ sharing the same eigenvalues as $L_z$ in the sense that 
\begin{equation}
    L_j = M^\dag L_z M,
\end{equation}
for $M$ being the unitary matrix diagonalizing $L_j$. Then, if $L_z$ is diagonal $(L_z)_{\alpha \beta} \propto \delta_{\alpha, \beta}$, the Jordan-Schwinger operator corresponding to $L_j$ can be written as
\begin{equation}
    \boldsymbol{L}_j = \boldsymbol{a}_\alpha^\dag M^\dag_{\alpha\gamma}\left(L_z\right)_{\gamma\gamma}M_{\gamma\beta} \boldsymbol{a}_\beta = \left(L_z\right)_{\gamma\gamma}\boldsymbol{N}'_\gamma,
\end{equation}
where $\boldsymbol{N}'_\gamma$ specifies the occupation numbers associated with transformed annihilation operators $\boldsymbol{a}'_\gamma = M_{\gamma\beta}\boldsymbol{a}_\beta$ (analogously for the creation operators). Since this transformation preserves the total particle number (as $M$ is unitary), it corresponds to a passive transformation in the language of quantum optics and allows us to construct a relation between the spin basis with respect to $\boldsymbol{L}_j$ and the Fock basis in two steps. First, we relate eigenstates of the CSCO including the spin component of interest $\boldsymbol{L}_j$ to the transformed Fock states $\{ \ket{N'_1}, \dots, \ket{N'_n} \}$ (with respect to which $\boldsymbol{L}_i$ is diagonal) following \autoref{subsec:Algorithm}, namely
\begin{equation}
    \ket{N,l',l_j,C'}=\sum_{N'_1,\dots,N'_n} c^{N,l',l_j,C'}_{N'_1,\dots,N'_n} \ket{N'_1,\dots,N'_n},
\end{equation}
for some coefficients $c$. In the second step, we transform $\ket{N'_1,\dots,N'_n}$ to the original mode operators using the matrix $M$. Note here that neither the total spin $l$ nor the counting number $C$ are conserved when changing between $\boldsymbol{L}_z$ and $\boldsymbol{L}_j$ eigenstates in general, as $\boldsymbol{L}_j$ may even belong to a different representation of $\mathfrak{su}(2)$. 

In the special case where $L_j$ and $L_z$ form a common representation of $\mathfrak{su}(2)$, the transformation $M$ reduces to a $n \times n$ rotation matrix $M \in SO(3)$ generated by $L_j,L_z$, and their corresponding third $\mathfrak{su}(2)$ matrix. Then, $L_j$ describes the angular momentum with respect to some axes on the same sphere as $L_z$. As the Jordan-Schwinger map conserves the Lie bracket, these properties extend to the spin operators $\boldsymbol{L}_j$ and $\boldsymbol{L}_z$, but also to the rotation operator $\boldsymbol{M}$, which is defined as the exponentiation of the former three (note that this does \textit{not} coincide with the image of the Jordan-Schwinger map applied to $M$). We parameterize the rotation in terms of the three Euler angles $\alpha,\beta,\gamma \in [0, 2\pi)$ such that the corresponding operator reads
\begin{equation}
    \boldsymbol{M} (\alpha, \beta, \gamma) = e^{- i \alpha \boldsymbol{L}_z} \, e^{- i \beta \boldsymbol{L}_y} \, e^{- i \gamma \boldsymbol{L}_z}.
\end{equation}
Then, the eigenstates of $\boldsymbol{L}_j$ and $\boldsymbol{L}_z$, are, for a fixed eigenvalue $l_z = l_j$, related via \cite{Sakurai2020}
\begin{equation}
    \begin{split}
        &\ket{N,l,l_j,C}_j \\
        &= \boldsymbol{M} (\alpha,\beta,\gamma )\ket{N,l,l_z,C}_z \\
        &= \sum_{l'_z} e^{-i\alpha l'_z} \, d^l_{l'_z l_j} (\beta) \, e^{-i\gamma l_j} \ket{N,l,l'_z,C}_z,
    \end{split}
    \label{eq:BasisRotation}
\end{equation}
thereby preserving all other quantum numbers $N, l, C$, where $d^l_{l'_z l_j} (\beta)$ denotes the real Wigner small $d$ matrix
\begin{equation}
    d^l_{l'_z l_j} (\beta) = \braket{N, l, l'_z, C | e^{- i \beta \boldsymbol{J}_y} | N, l, l_j, C}.
\end{equation}
When $\boldsymbol{L}_j = \boldsymbol{L}_x$ or $\boldsymbol{L}_y$, Eq. \eqref{eq:BasisRotation} reduces to
\begin{equation}
    \begin{split}
        \ket{N,l,l_x,C}_x &= \sum_{l'_z} d^l_{l'_z l_x}\left(\frac{\pi}{2}\right) e^{i\frac{\pi}{2} l'_z} \ket{N,l,l'_z,C}_z, \\ 
        \ket{N,l,l_y,C}_y &= \sum_{l'_z} e^{-i\frac{\pi}{2} l'_z} d^l_{l'_z l_y}\left(- \frac{\pi}{2}\right)\ket{N,l,l'_z,C}_z,
    \end{split}
    \label{eq:BasisRotationxy}
\end{equation}
respectively. Note here that Eqs. \eqref{eq:BasisRotation} and \eqref{eq:BasisRotationxy} hold independently of the considered representation in the sense that spin operators corresponding to the same representation of $\mathfrak{su}(2)$ are always related by rotations. However, the chosen representation determines the ladder operator $\boldsymbol{L}_-$ and, therefore, leads to differences when constructing their spin bases and defining the above rotations.

\section{From three modes to spins}
\label{sec:ThreeModes}

\subsection{Setup}
\label{subsec:Setup}

\begin{figure*}
    \centering
    \includegraphics[width=0.7\linewidth]{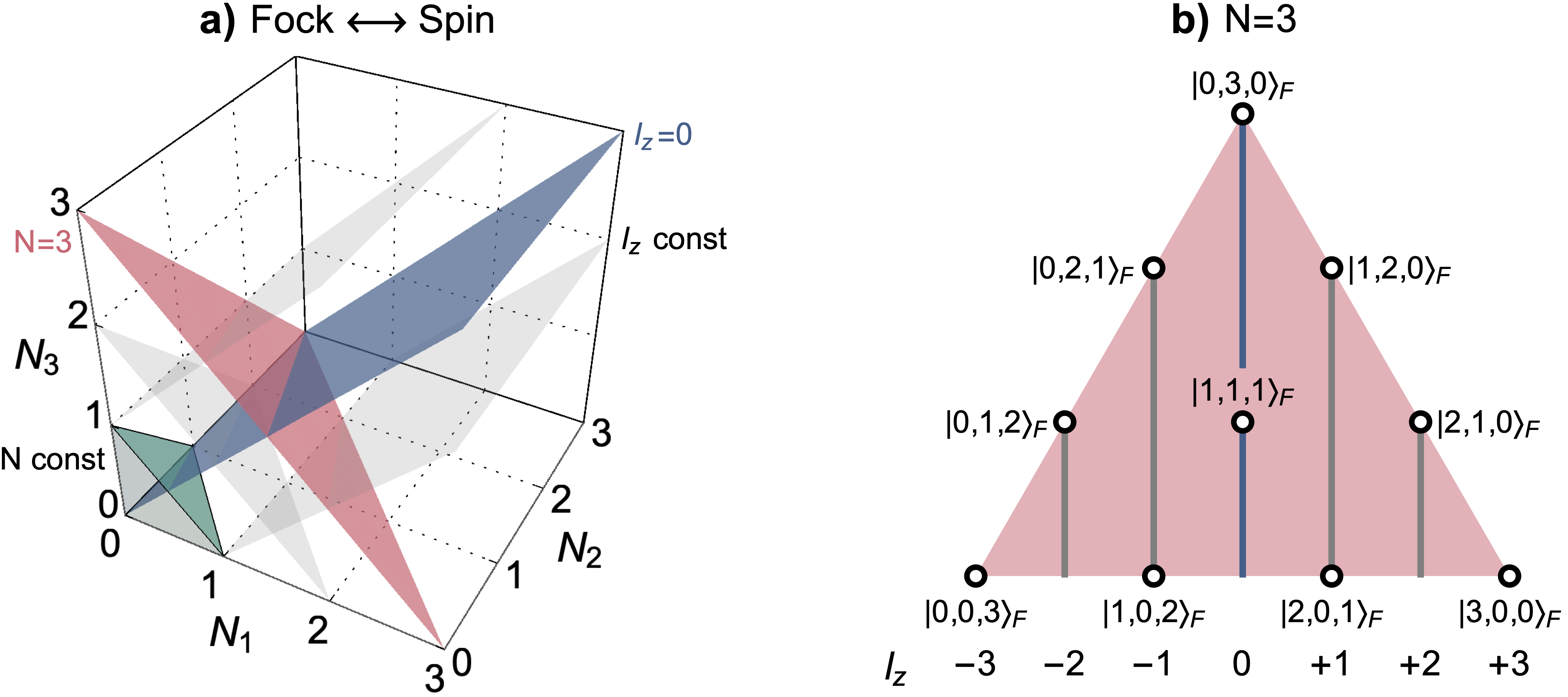}
    \caption{\textbf{a)} Graphical representation of the three-mode Fock space, which corresponds to the cubic lattice $(N_1, N_2, N_3)$, with the constraint $N_j \le 1$ for fermions (green shaded region). As discussed in \autoref{sec:Construction}, spin states are constructed along intersections of the spaces corresponding to fixed particle numbers, that is, simplexes defined by $N=N_1+ N_2 + N_3$, see red simplex for $N=3$, and fixed $z$-component, \textit{i.e.}, planes along $l_z = N_3 - N_1$, see blue plane for $l_z=0$. \textbf{b)} Simplex of $N=3$ particles with lines of constant $l_z$. For bosons, the three particles can be freely distributed over the three modes, which results in one (two) possible configuration(s) each for $l_z = \pm 2, \pm 3 \, (0, \pm 1)$, cf. \autoref{fig:Degeneracy} \textbf{a)}. The only admissible three-particle fermionic state is the central state $\ket{1,1,1}_F$.}
    \label{fig:ThreeModes}
\end{figure*}

\subsubsection{Three-mode spin operators}
As an application, we consider a three-mode ($n=3$) system and \textit{explicitly} construct a complete set of spin states along the lines of \autoref{subsec:Algorithm}. The representation in which $L_z$ is diagonal is the adjoint representation of $\mathfrak{su}(2)$\footnote{This corresponds to the fundamental representation of $\mathfrak{so}(3)$ up to a similarity transform.}. Using the expressions in \autoref{subsec:BeyondTwoModes}, it consists of the spin operators
\begin{equation}
    \begin{split}
        \boldsymbol{L}_x &\equiv \frac{1}{\sqrt{2}} \left[ \left(\boldsymbol{a}_1^\dag + \boldsymbol{a}_3^\dag \right) \boldsymbol{a}_2 + \boldsymbol{a}^\dag_2 \left(\boldsymbol{a}_1 + \boldsymbol{a}_3 \right) \right], \\
        \boldsymbol{L}_y &\equiv \frac{i}{\sqrt{2}} \left[ \left( - \boldsymbol{a}_1^\dag + \boldsymbol{a}_3^\dag \right) \boldsymbol{a}_2 + \boldsymbol{a}^\dag_2 \left(\boldsymbol{a}_1 - \boldsymbol{a}_3 \right) \right], \\
        \boldsymbol{L}_z &\equiv \boldsymbol{a}_1^\dag \boldsymbol{a}_1 - \boldsymbol{a}^\dag_3 \boldsymbol{a}_3,
    \end{split}
    \label{eq:SpinOperatorsThreeModes}
\end{equation}
with ladder operators
\begin{equation}
    \boldsymbol{L}_{+} = \boldsymbol{a}_1^\dag \boldsymbol{a}_2 + \boldsymbol{a}_2^\dag \boldsymbol{a}_3, \quad \boldsymbol{L}_{-} = \boldsymbol{a}_1 \boldsymbol{a}_2^\dag + \boldsymbol{a}_2 \boldsymbol{a}_3^\dag,
    \label{eq:ThreeModeLadderOperators}
\end{equation}
and total spin
\begin{equation}
    \begin{split}
        \boldsymbol{L}^2 &= \boldsymbol{N}_1 \left( \mathds{1} + \boldsymbol{N}_1 \pm 2 \boldsymbol{N}_2 - \boldsymbol{N}_3 \right) \\
        &\hspace{0.4cm}+ \boldsymbol{N}_3 \left(3 \cdot \mathds{1} - \boldsymbol{N}_1 \pm 2 \boldsymbol{N}_2 + \boldsymbol{N}_3 \right) \\ 
        &\hspace{0.4cm}+2 \left[ \boldsymbol{a}_1 (\boldsymbol{a}_2^\dag)^2 + \boldsymbol{a}_1^\dag (\boldsymbol{a}_2)^2 \right] \boldsymbol{a}_3,
    \end{split}
\end{equation}
where $+$ or $-$ for bosons or fermions, respectively (note that the last term is absent in the fermionic case).

Since $n$ is odd, the two spin quantum numbers $l,l_z$ attain integer values for arbitrary particle numbers $N$. For bosons, they are drawn from $l,l_z \in \{-N, \dots, N \}$, while for fermions we find $l,l_z \in \{-1, 0, 1\}$ for $N=1,2$ and $l=l_z=0$ for $N=3$. In analogy to \autoref{fig:TwoModes} \textbf{a)}, we depict the three-mode Fock space as a cubic lattice, see \autoref{fig:ThreeModes} \textbf{a)}. In this representation, a fixed total particle number $N = N_1 + N_2 + N_3$ corresponds to a simplex which intersects with every coordinate axis at a distance $N$ from the origin, see, \textit{e.g.}, the red simplex representing $N=3$ bosonic particles. Note here that Pauli's principle strongly restricts the fermionic Fock space, see green simplex. In contrast, a fixed $z$-component $l_z = N_1 - N_3$ is described by a plane, see, \textit{e.g.}, the blue plane for $l_z=0$. The degeneracy function counts the number of Fock states along an intersection of these two spaces, which, for $n=3$ modes, reads [see \autoref{fig:Degeneracy} \textbf{a)}]
\begin{equation}
    \mathfrak{g}(3,N,l_z) = \begin{cases}
        1 + \floor{\frac{N-\abs{l_z}}{2}} & \text{bosons}, \\
        1 & \text{fermions}.
    \end{cases}
    \label{eq:DegeneracyFunctionThreeModes}
\end{equation}

\subsubsection{Basis relations for up to $N=3$ particles}
We now discuss the basis relation between Fock and spin states for up to $N=3$ particles of arbitrary type, which we also tabularize in \hyperref[app:ThreeModes]{Appendix B}. We start from the trivial case $N=0$, for which $\ket{0,0,0}_s = \ket{0,0,0}_F$, for both bosons and fermions (we omit the counting number $C$ for brevity as it is irrelevant for three modes). For a single particle $N=1$, the total spin is always $l=1$, with the particle nature still being irrelevant and the $z$-component $l_z$ being fully determined by the populated mode. More precisely, $l_z=-1,0,1$ when the particle is in the last, middle, and first mode, respectively. 

The first non-trivial case is $N=2$, for which bosons and fermions behave rather differently. While two bosons can represent a total spin of $l=2$, two fermions still correspond to $l=1$. This can be understood in analogy to Dirac's sea: the absence of one fermion may be regarded as a hole in a sea of fermions and, in particular, as a fermion itself. Hence, one fermion and two holes, \textit{i.e.}, $N=1$, is equivalent to two fermions and one hole, \textit{i.e.}, $N=2$, when exchanging $\ket{0}$ with $\ket{1}$ in each mode. Further, we remark that for $N=2$ bosons, a vanishing $z$-component $l_z=0$ is attained by the two Fock states $\ket{1,0,1}_F$ and $\ket{0,2,0}_F$, see also petrol curve in \autoref{fig:Degeneracy} \textbf{a)} at $l_z=0$. 

At last, we consider $N=3$ particles, see \autoref{fig:ThreeModes} \textbf{b)}. For fermions, all modes have to be populated once, which maps onto the spin-zero state, \textit{i.e.}, $\ket{3,0,0}_s = \ket{1,1,1}_F$. Note here the equivalence to the $N=0$ case in the sense that three particles and zero holes are equivalent to zero particles and three holes. In contrast, a variety of spin states arises for bosons: pairs of spin states with $l_z = 0, \pm 1$ (distinguished by $l=1,3$) are mapped onto pairs of Fock states, see, \textit{e.g.}, the blue line in \autoref{fig:ThreeModes} \textbf{b)} corresponding to $l_z=0$, which represents the intersection of the $N=3$ and $l_z=0$ spaces in \autoref{fig:ThreeModes} \textbf{a)}. For $l_z =\pm 2, \pm 3$, spin and Fock states exhibit one-to-one mappings.

\subsection{Explicit basis relations for $N$ bosons}
\label{subsec:BasisRelations}
While no fermionic states can exist beyond $N=3$ particles, the bosonic particle number $N$ is unconstrained in general. Remarkably, we still find explicit relations between spin and Fock states for arbitrary $N$. The basic idea is to apply the lowering operator to the kernel of the raising operator on the space where the $z$-component is maximal, \textit{i.e.}, $\ket{N,l,l_z}_s \propto \boldsymbol{L}_-^{l-l_z} [\ker (\boldsymbol{L}_+) \cup \mathcal{E}_{N,l}]$.

Starting from any normalized pure state $\ket{\psi}$ with quantum numbers $N, l, l_z$ where $l_z \ge 0$, Eq. \eqref{eq:DegeneracyFunctionThreeModes} shows that it can always be written as
\begin{equation}
    \ket{\psi} = \sum_{j=0}^{\floor{\frac{N-l_z}{2}}} \lambda (N,l,l_z,j) \ket{l_z+j,N-l_z-2j,j}_F,
\end{equation}
for some coefficients $\lambda (N,l,l_z,j)$. Therefore, we have that
\begin{equation}
    \ket{N,l,l}_s = \sum_{j=0}^{\frac{N-l}{2}} \lambda_1 (N,l,j) \ket{l+j,N-l-2j,j}_F,
    \label{eq:ThreeModeHighestWeightState}
\end{equation}
for any $l$ having the same parity as $N$, a requirement which is implied by the three-mode degeneracy function \eqref{eq:DegeneracyFunctionThreeModes}, see also \autoref{fig:Degeneracy} \textbf{a)}. We compute the corresponding coefficients $\lambda_1 (N,l,j)$ by acting with the raising operator $\boldsymbol{L}_+$ and noting that the result has to vanish, whereafter we impose normalization, see \hyperref[app:ThreeModeCoefficients]{Appendix C} for explicit expressions.

To obtain other values for the $z$-component $l_z < l$, we instead apply the lowering operator as discussed in \autoref{subsec:Algorithm}, which yields
\begin{equation}
    \ket{N,l,l_z}_s = \lambda_2 (l,l_z) \boldsymbol{L}_-^{l-l_z} \ket{N,l,l}_s.
\end{equation}
Therein, the prefactors 
\begin{equation}
    \lambda_2 (l,l_z) = \sqrt{\frac{2^{l-l_z} (l+l_z)!}{(2l)!(l-l_z)!}}
    \label{eq:ThreeModeCoefficients2}
\end{equation}
follow from computing the action of the lowering operator $\boldsymbol{L}_-$ on spin states, that is, $\boldsymbol{L}_- \ket{N,l,l_z}_s = \sqrt{[l(l+1)-l_z (l_z-1)]/2} \ket{N,l,l_z-1}_s$. Note here that the resulting state is already normalized as $\lambda_1$ has been normalized before. Then, we employ a formula for integer powers of the lowering operator
\begin{equation}
    \begin{split}
        \boldsymbol{L}_-^d = \sum_{e=0}^{\floor{\frac{d}{2}}}\sum_{f=0}^{d-2e} \lambda_3 (d,e,f) \boldsymbol{a}_1^{d-e-f} \left(\boldsymbol{a}_2^\dag\right)^{d-2e-f}\boldsymbol{a}_2^{f} \left(\boldsymbol{a}_3^\dag\right)^{e+f},
    \end{split}
    \label{eq:ThreeModeLoweringOperatorPowers}
\end{equation}
with coefficients
\begin{equation}
    \lambda_3 (d,e,f)=\frac{d!}{2^e e! f! (d-2e-f)!},
    \label{eq:ThreeModeCoefficients3}
\end{equation}
which we prove in \hyperref[app:ThreeModeLoweringOperator]{Appendix D}. After applying the remaining mode operators, which generates yet another set of coefficients $\lambda_4 (N,l,l_z,j,e,f)$, we finally find
\begin{equation}
    \begin{split}
        &\ket{N,l,l_z}_s = \sum_{j=0}^{\frac{N-l}{2}} \sum_{e=0}^{\floor{\frac{l-l_z}{2}}} \sum_{f=0}^{l-l_z-2e} \lambda (N,l,l_z,j,e,f) \\
        &\ket{l_z+j+e+f,N-l_z-2j-2e-2f,j+e+f}_F,
    \end{split}
    \label{eq:ThreeModeSpinState}
\end{equation}
where the global coefficient decomposes as $\lambda = \prod_{i=1}^4 \lambda_i$, see \hyperref[app:ThreeModeCoefficients2]{Appendix E} for explicit expressions.

\subsection{Tripartite entanglement in the spin picture}
\label{subsec:TripartiteEntanglement}

\subsubsection{Generalized GHZ and W states}
With \eqref{eq:ThreeModeSpinState} at hand, we present a fresh view of tripartite entanglement in terms of spins. To this end, we introduce the two classes of genuine tripartite entangled states, generalized to $N \ge 1$ particles \textit{per mode} (hence, each party has one mode and the total particle number is $3N$ in the following): the W state \cite{Duer2000}
\begin{equation}
    \ket{\psi_\text{W}} = \frac{1}{\sqrt{3}} \left( \ket{N,0,0}_F + \ket{0,N,0}_F + \ket{0,0,N}_F \right),
    \label{eq:WState}
\end{equation}
and the GHZ state \cite{Greenberger1989}
\begin{equation}
    \ket{\psi_\text{GHZ}} = \frac{1}{\sqrt{2}} \left( \ket{0,0,0}_F + \ket{N,N,N}_F \right),
    \label{eq:GHZState}
\end{equation}
where $N \in \mathbb{N}_+$ for bosons and $N=1$ for fermions (note that we have omitted possible phases). The two states \eqref{eq:WState} and \eqref{eq:GHZState} constitute non-equivalent classes of tripartite entanglement in the sense that one state cannot be transformed into the other by Local Operations and Classical Communication (LOCCs). Interestingly, a generic pure state of three qubits can be transformed to a GHZ state using LOCC with a finite probability, showing that the class of W states is of zero measure in this case. Further, the two states behave differently when one mode is traced out: the GHZ state turns into a separable mixture, while the reduced W state remains entangled.

\begin{figure*}
    \centering
    \includegraphics[width=0.99\linewidth]{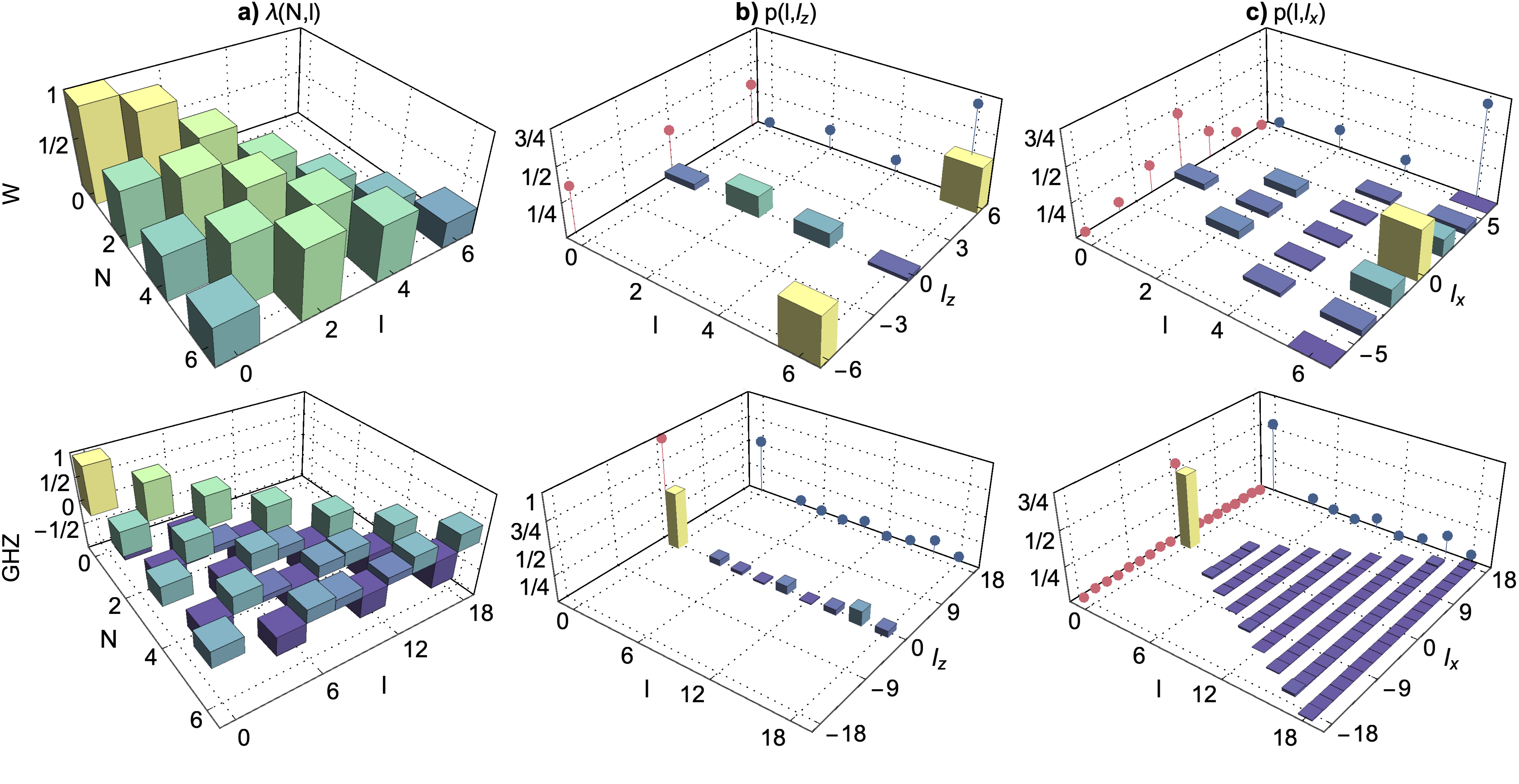}
    \caption{\textbf{a)} Coefficients $\lambda(N,l)$ representing the overlap between the W (upper row) or GHZ state (lower row) and the $l_z=0$ spin states. \textbf{b)} and \textbf{c}) Spin measurement distributions in the $(l,l_z)$ and $(l,l_x)$ bases together with their marginal distributions over $l$ (blue) and $l_z$ and $l_x$ (red), respectively, for $N=6$ particles per mode (in all cases, only non-zero values are shown). }
    \label{fig:SpinMeasurements}
\end{figure*}

To understand the differences between W and GHZ states in the spin picture, we first note that two of the three terms required for the W state correspond to the highest and lowest weight states for $N$ particles by definition, \textit{i.e.}, $\ket{N,0,0}_F = \ket{N,N,N}_s$ and $\ket{0,0,N}_F = \ket{N,N,-N}_s$, respectively, in which case the total spin, as well as the absolute value of the $z$-component, agree with the mode-wise particle number $N=l=\abs{l_z}$. Second, we remark that the two remaining terms $\ket{0,N,0}_F$ and $\ket{N,N,N}_F$ both correspond to spin states characterized by a vanishing $z$-component $l_z=0$. However, they differ in total particle number (hence also in total spin $l$): while the former contains $N$ particles in total, the latter constitutes a $3N$-particle state. Since the total spin $l$ is not fixed in either of the two cases, the two states admit series expansions of the form \eqref{eq:ThreeModeSpinState}, namely
\begin{equation}
    \begin{split}
        \ket{\psi_{\text{W}}} &= \frac{1}{\sqrt{3}} \Bigg( \ket{N,N,N}_s + \ket{N,N,-N}_s \\
        &\hspace{1.25cm}+ \sum_{l=0,1; +2}^{N} \lambda_{\text{W}} (N,l) \ket{N,l,0}_s \Bigg), \\
        \ket{\psi_{\text{GHZ}}} &= \frac{1}{\sqrt{2}} \Bigg( \ket{0,0,0}_s \\
        &\hspace{1.25cm}+ \sum_{l=0,1; +2}^{3N} \lambda_{\text{GHZ}} (N,l) \ket{3N,l,0}_s \Bigg), 
    \end{split}
    \label{eq:GHZWStatesSpinBasis}
\end{equation}
with the sums over $l$ running from zero or one if $N$ is even or odd, respectively, in steps of two. The first of the two remaining coefficients, $\lambda_{\text{W}}$, can be evaluated analytically (see \hyperref[app:ThreeModeCoefficientsGHZW]{Appendix F} for details), while the second, $\lambda_{\text{GHZ}}$, requires numerical summation. They read
\begin{equation}
    \begin{split}
        \lambda_{\text{W}} (N,l) &= \lambda (N,l,0,0,0,0) \\
        &= \sqrt{\frac{N!(2 l+1)}{(N-l)!!(N+l+1)!!}}, \\
        \lambda_{\text{GHZ}} (N,l) &= \sum_{j=0}^{\frac{3N-l}{2}} \sum_{e=0}^{\floor{\frac{l}{2}}} \lambda (3N,l,0,j,e,N-j-e),
    \end{split}
    \label{eq:ThreeModeCoefficientsGHZW}
\end{equation}
and are plotted in \autoref{fig:SpinMeasurements} \textbf{a)}.

\subsubsection{Spin measurement distributions}
We are now ready to compute the outcome distributions associated with spin measurements -- a task for which the basis relations between spin and Fock states are crucial. The joint probability distributions in the $z$- and $x$-bases are defined as [see also Eq. \eqref{eq:BasisRotationxy}]
\begin{equation}
    \begin{split}
        p(l,l_z) &= \sum_{N'=0}^{\infty} \abs{\braket{N',l,l_z | \psi}}^2, \\
        p(l,l_x) &= \sum_{N'=0}^{\infty} \Big\lvert \sum_{l_z} d_{l_z l_x}^{l} e^{-i\frac{\pi}{2}l_z} \left( \frac{\pi}{2} \right) \braket{N',l,l_z | \psi} \Big\lvert^2,
    \end{split}
    \label{eq:JointProbabilityDistributions}
\end{equation}
respectively. The $y$-distribution follows analogously but is omitted since $p(l,l_x)=p(l,l_y=l_x)$ for both states of interest. The marginal distributions are obtained via
\begin{equation}
        p(l) = \sum_{l_j = -l}^l p(l,l_j), \quad p(l_j) = \sum_{l=n;-2}^{\abs{l_j}} p(l,l_j).
\end{equation}

We start with the joint distributions. Using \eqref{eq:JointProbabilityDistributions} together with \eqref{eq:GHZWStatesSpinBasis} we find
\begin{equation}
    \begin{split}
        p_{\text{W}}(l,l_z) &= \frac{1}{3} \left[ \delta_{l N} (\delta_{l_z N} + \delta_{l_z -N}) + \delta_{l_z 0} \lambda^2_{\text{W}} (N,l) \right], \\
        p_{\text{GHZ}}(l,l_z) &= \frac{1}{2} \delta_{l_z 0} \left[ \delta_{l 0} + \lambda^2_{\text{GHZ}} (N,l) \right],
    \end{split}
\end{equation}
and
\begin{equation}
    \begin{split}
        p_{\text{W}}(l,l_x) &= \frac{1}{3} \Big \lvert \delta_{l N} i^N \left[ (-1)^N d_{N l_x}^{N} \left( \frac{\pi}{2} \right) + d_{-N l_x}^{N} \left( \frac{\pi}{2} \right) \right] \\
        &\hspace{0.8cm}+ d_{0 l_x}^{l} \left( \frac{\pi}{2} \right) \lambda_{\text{W}} (N,l) \Big\lvert^2, \\
        p_{\text{GHZ}}(l,l_x) &= \frac{1}{2} \left\{ \delta_{l 0} \delta_{l_z 0} + \left[ d_{0 l_x}^l \left( \frac{\pi}{2} \right) \lambda_{\text{GHZ}} (N,l) \right]^2 \right\},
    \end{split}
\end{equation}
see \autoref{fig:SpinMeasurements} \textbf{b)} and \textbf{c)}, respectively. The total spin marginals read (see blue distributions)
\begin{equation}
    \begin{split}
        p_{\text{W}}(l) &= \frac{1}{3} \left[ 2 \delta_{l N} + \lambda^2_{\text{W}} (N,l) \right], \\
        p_{\text{GHZ}}(l) &= \frac{1}{2} \left[ \delta_{l 0} + \lambda^2_{\text{GHZ}} (N,l) \right], \\
    \end{split}
\end{equation}
while for the $z$-components we obtain (see red distributions)
\begin{equation}
    \begin{split}
        p_{\text{W}}(l_z) &= \frac{1}{3} \left(\delta_{l_z N} + \delta_{l_z -N} + \delta_{l_z 0} \right), \\ 
        p_{\text{GHZ}}(l_z) &= \delta_{l_z 0}.
    \end{split}
\end{equation}
The $x$-distributions admit no closed forms in general.

Interestingly, we find that the two classes of states are in several ways complementary in the spin picture. Most importantly, the GHZ state [see lower row of \autoref{fig:SpinMeasurements} \textbf{b)} and \textbf{c)}] assigns non-zero probabilities to total spin measurements up to the global maximum set by the total particle number $l \le 3N$. In contrast, the W state [see upper row of \autoref{fig:SpinMeasurements} \textbf{b)} and \textbf{c)}] is constrained by $p(l,l_j) = 0$ for $l,l_j > N$, and thus lives in a much smaller space. This is reminiscent and may be connected to the fact that the GHZ class represents generic pure states up to a LOCC, while the W class is of zero measure.

Also, the shapes of the measurement distributions strongly differ: First, the W state assigns equal probabilities to the three extremal values $l_z=\pm l, 0$. In contrast, the GHZ state gives a contribution only for $l_z = 0$, showing that the W state's $l_z$ distribution is majorized by the GHZ state's one, \textit{i.e.}, $p_{\text{W}} (l_z) \prec p_{\text{GHZ}} (l_z)$\footnote{Here, we extended the domain of $l_z$ for the W state to $l_z \in [-3N, 3N]$ and set $p_{\text{W}} (l_z) = 0$ for $\abs{l_z}>N$}. We observe a similar behavior, although less pronounced, for the $l_x$ distributions and, therefore, also for the $l_y$ distributions. Second, the total spin distributions $p(l)$ are somewhat reversed. For the W state, $l=N$ exhibits the largest probability, which is the maximal value as discussed above. In contrast, for the GHZ state, we observe that $l=0$ possesses the highest probability. Third, we find $p_{\text{GHZ}} (l) = 0$ for $l=2$, which results from $\lambda_{\text{GHZ}} (N,2)=0$, while no such constraint appears for the W state.

\section{Discussion} 
\label{sec:Conclusion}
In conclusion, we have proposed an extension of the Jordan-Schwinger map that allows one to describe multi-mode quantum systems of both bosonic and fermionic types in terms of spins. While our construction of a complete spin basis is only algorithmic, in general, we found explicit relations between spin and Fock states for the three-mode case, with the latter being directly applicable to all kinds of three-level systems. 

Our results find direct applications in quantum optics: it has been of recent interest to devise simple measurement schemes, that is, passive optical circuits followed by photon-number detection, for various two- or three-mode spin observables designed to reveal quantum features of light \cite{Agarwal2005,Hertz2019b,Arnhem2022,Griffet2023a,Griffet2023b}. So far, analyses of the corresponding outcome distributions have been limited to low-order moments only, as their evaluation remains feasible in the mode picture. Here, our precise relations between spin and Fock states pave the way for a complete characterization, allowing one to probe the full statistics of spin measurements, also using functional quantities such as the Shannon entropy. In this regard, it is also an interesting problem to construct sets of optical circuits for reading out general classes of spin operators.

More generally, besides computing similar relations for four- or five-mode systems, we consider investigating complex quantum many-body systems in the spin instead of mode picture an essential venue for future work. What we have in mind here are spin-1 Bose-Einstein condensates, where engineerable Hamiltonians can be expressed in terms of spin operators constructed via the Jordan-Schwinger map. The conventional approach to solving such theories, \textit{i.e.}, finding the Hamiltonian's ground states, requires lengthy calculations solely carried out in the mode picture. Instead, our mapping allows one to solve the model in the much simpler spin basis and transform back to Fock space in a single step, indicating that the complexity of the problem could be entirely shifted to the now-existing transformation between Fock and spin states. This logic may also be beneficial for other many-body theories, such as the Bose-Hubbard model, or for analyzing simple three-mode systems, as we have exemplified by characterizing genuine tripartite entangled states.

A central outlook question regards the generality of our findings. In this work, we focused on spin operators as representations of $\mathfrak{su}(2)$, but the Jordan-Schwinger map is neither limited to this algebra nor to spin operators. Our spin-basis construction ultimately relies on identifying the action of lowering and raising operators of a given algebra on multi-mode Fock states, whose existence is guaranteed for every semi-simple Lie algebra. Therefore, we expect our method to be extendable to other algebras of physical interest such as, \textit{e.g.}, $\mathfrak{su}(3)$. One may also consider operators that do not belong to representations of $\mathfrak{su}(2)$ (or other algebras). For instance, starting from any hermitian diagonal matrix, the corresponding Jordan-Schwinger operator commutes with the total number operator and, therefore, can be used to construct a CSCO.

On the mathematical side, we consider the link between the degeneracy function and Gaussian polynomials an exciting starting point for future work. While we discovered such a link for bosonic degrees of freedom, finding its fermionic counterpart to another polynomial series remains an open problem. In this context, we also regard an analysis of the continuum limit $n \to \infty$ of particular importance, upon which our findings could be extended from multiple modes to quantum field theories. 

\section*{Acknowledgements} We thank Célia Griffet for initial discussions on the subject. B. D. is a Research Fellow of the F.R.S.- FNRS. T. H. and N. J. C. acknowledge support from the European Union under project ShoQC within the ERA-NET Cofund in Quantum Technologies (QuantERA) program and from the F.R.S.- FNRS under project CHEQS within the Excellence of Science (EOS) program.


\appendix

\section{From Gaussian polynomials to the degeneracy function}
\label{app:GaussianPolynomials}
We prove \eqref{eq:GaussianPolynomialsDegeneracyRelation} by demonstrating that the recursion relation \eqref{eq:GaussianPolynomialsRecursionRelation} with initial conditions \eqref{eq:GaussianPolynomialsInitialConditions} is fulfilled. Starting from the right-hand side of \eqref{eq:GaussianPolynomialsRecursionRelation}, we find 
\begin{equation}
    \begin{split}
        &\begin{bmatrix}
            j-1 \\ k-1
        \end{bmatrix}_q + q^k
        \begin{bmatrix}
            j-1 \\ k
        \end{bmatrix}_q \\
        &= \sum_{i=0}^{(k - 1) (j-k)} \mathfrak{h} (j-k+1, k-1, i+k-1) \, q^i \\
        &\hspace{0.4cm}+ \sum_{i=k}^{k (j-k)} \mathfrak{h} (j-k, k, i) \, q^i.
    \end{split}
\end{equation}
Then, we note that both summation limits can be extended: for the first term, the maximum score is $(k-1)(j-k+1)$, and thus $\mathfrak{h} (j-k+1, k-1, i+k-1) = 0$ if $i > (k-1) (j-k)$. Similarly, the minimum score for the second term is $k$, and hence $\mathfrak{h} (j-k,k,i) = 0$ when $i < k$. Combining the two sums and using the recursion relation \eqref{eq:hDefinition} for the degeneracy function results in
\begin{equation}
    \begin{split}
        &\begin{bmatrix}
            j-1 \\ k-1
        \end{bmatrix}_q + q^k
        \begin{bmatrix}
            j-1 \\ k
        \end{bmatrix}_q \\
        &= \sum_{i=0}^{k (j-k)} \Big[ \mathfrak{h} (j-k+1, k-1, i+k-1) \\
        &\hspace{1.5cm}+\mathfrak{h} (j-k, k, i) \Big] \, q^i  \\
        &= \sum_{i=0}^{k (j-k)} \mathfrak{h} (j-k+1, k, i+k) \, q^i \\
        &= \begin{bmatrix}
            j \\ k
        \end{bmatrix}_q,
    \end{split}
\end{equation}
as desired. The equivalence of the initial conditions follows from verifying \eqref{eq:GaussianPolynomialsInitialConditions} using \eqref{eq:BosonicInitialConditions}, to wit
\begin{equation}
    \begin{bmatrix}
        j \\ 
    \end{bmatrix}_q
    = \mathfrak{h} (1, j, j) = 1, \quad \begin{bmatrix}
        j \\ 0
    \end{bmatrix}_q
    = \mathfrak{h} (j+1, 0, 0) = 1,
\end{equation}
which concludes the proof.

\section{Basis relations for $n=3$ modes and $N \le 3$ particles}
\label{app:ThreeModes}
We summarize the relations between spin states and three-mode Fock states for up to three particles in \autoref{tab:BasisRelationsBosons} and \autoref{tab:BasisRelationsFermions} for bosons and fermions, respectively.

\renewcommand{\arraystretch}{1.6}
\setlength{\tabcolsep}{1.4pt}
\begin{table*}[t!]
  \centering
    \begin{tabular}{c  l l  l l  l l  l l}
        \toprule
        \multirow{2}{*}{\, $\boldsymbol{l_z}$ \,} & \multicolumn{2}{c}{$\boldsymbol{N=0}$} & \multicolumn{2}{c}{$\boldsymbol{N=1}$} & \multicolumn{2}{c}{$\boldsymbol{N=2}$} & \multicolumn{2}{c}{$\boldsymbol{N=3}$} \\
         & Spins & Bosons & Spins & Bosons & Spins & Bosons & Spins & Bosons \\
        \midrule
          +3 & & & & & & & $\ket{3,3,3}_s$ & $\ket{3,0,0}_F$ \\
          +2 & & & & & $\ket{2,2,2}_s$ & $\ket{2,0,0}_F$ & $\ket{3,3,2}_s$ & $\ket{2,1,0}_F$ \\
          \multirow{2}{*}{+1} & & & \multirow{2}{*}{$\ket{1,1,1}_s$} & \multirow{2}{*}{$\ket{1,0,0}_F$} & \multirow{2}{*}{$\ket{2,2,1}_s$} & \multirow{2}{*}{$\ket{1,1,0}_F$} & $\ket{3,3,1}_s$ & $\frac{1}{\sqrt{5}} \left( 2 \ket{1,2,0}_F + \ket{2,0,1}_F \right)$ \\
           & & & & & & & $\ket{3,1,1}_s$ & $\frac{1}{\sqrt{5}} \left( \ket{1,2,0}_F - 2 \ket{2,0,1}_F \right)$ \\
          \multirow{2}{*}{0} & \multirow{2}{*}{$\ket{0,0,0}_s$} & \multirow{2}{*}{$\ket{0,0,0}_F$} & \multirow{2}{*}{$\ket{1,1,0}_s$} & \multirow{2}{*}{$\ket{0,1,0}_F$} & $\ket{2,2,0}_s$ & $\frac{1}{\sqrt{3}} \left( \sqrt{2} \ket{0,2,0}_F + \ket{1,0,1}_F \right)$  & $\ket{3,3,0}_s$ & $\frac{1}{\sqrt{5}} \left( \sqrt{2} \ket{0,3,0}_F + \sqrt{3} \ket{1,1,1}_F \right)$ \\
           & & & & & $\ket{2,0,0}_s$ & $\frac{1}{\sqrt{3}} \left(\ket{0,2,0}_F - \sqrt{2} \ket{1,0,1}_F \right)$ & $\ket{3,1,0}_s$ & $\frac{1}{\sqrt{5}} \left( \sqrt{3} \ket{0,3,0}_F - \sqrt{2} \ket{1,1,1}_F \right)$ \\
          \multirow{2}{*}{-1} & & & \multirow{2}{*}{$\ket{1,1,-1}_s$} & \multirow{2}{*}{$\ket{0,0,1}_F$} & \multirow{2}{*}{$\ket{2,2,-1}_s$} & \multirow{2}{*}{$\ket{0,1,1}_F$} & $\ket{3,3,-1}_s $ & $\frac{1}{\sqrt{5}} \left( 2 \ket{0,2,1}_F + \ket{1,0,2}_F \right)$ \\
          & & & & & & & $\ket{3,1,-1}_s$ & $\frac{1}{\sqrt{5}} \left( \ket{0,2,1}_F - 2 \ket{1,0,2}_F \right)$ \\
          -2 & & & & & $\ket{2,2,-2}_s$ & $\ket{0,0,2}_F$ & $\ket{3,3,-2}_s$ & $\ket{0,1,2}_F$ \\
          -3 & & & & & & & $\ket{3,3,-3}_s$ & $\ket{0,0,3}_F$ \\
        \bottomrule
    \end{tabular}
    \caption{Basis relations for up to $N=3$ bosons in $n=3$ modes. For brevity, we omit the counting number $C$.}
    \label{tab:BasisRelationsBosons}
    \vspace{0.8cm}
    \setlength{\tabcolsep}{2pt}
    \begin{tabular}{c l l  l l  l l  l l}
        \toprule
        \multirow{2}{*}{\, $\boldsymbol{l_z}$ \,} & \multicolumn{2}{c}{$\boldsymbol{N=0}$} & \multicolumn{2}{c}{$\boldsymbol{N=1}$} & \multicolumn{2}{c}{$\boldsymbol{N=2}$} & \multicolumn{2}{c}{$\boldsymbol{N=3}$} \\
         & Spins & Fermions & Spins & Fermions & Spins & Fermions & Spins & Fermions \\
        \midrule
          +1 & & & $\ket{1,1,1}_s$ & $\ket{1,0,0}_F$ & $\ket{2,1,1}_s$ & $\ket{1,1,0}_F$ \\
          0 & $\ket{0,0,0}_s$ & $\ket{0,0,0}_F$ & $\ket{1,1,0}_s$ & $\ket{0,1,0}_F$ & $\ket{2,1,0}_s$ & $\ket{1,0,1}_F$ & $\ket{3,0,0}$ & $\ket{1,1,1}_F$ \\
          -1 & & & $\ket{1,1,-1}_s$ & $\ket{0,0,1}_F$ & $\ket{2,1,-1}_s$ & $\ket{0,1,1}_F$ \\
        \bottomrule
    \end{tabular}
    \caption{Basis relations for up to $N=3$ fermions in $n=3$ modes, with $C$ omitted.}
    \label{tab:BasisRelationsFermions}
\end{table*}

\section{Coefficients of three-mode highest weight spin states}
\label{app:ThreeModeCoefficients}
Since \eqref{eq:ThreeModeHighestWeightState} corresponds to the unique highest weight state for which $l_z = l$, it is annihilated by the raising operator, \textit{i.e.}, $\boldsymbol{L}_+ \ket{N,l,l}_s = 0$ for all $l$. Applying the raising operator's explicit form \eqref{eq:ThreeModeLadderOperators} to \eqref{eq:ThreeModeHighestWeightState} and noting that the prefactor of the $j=(N-l)/2$ term vanishes results in the condition
\begin{equation}
    \begin{split}
        0 &= \boldsymbol{L}_+ \ket{N,l,l}_s \\
        &= \sum_{j = 0}^{\frac{N-l}{2}-1} \Big[ \lambda_1 (N,l,j) \sqrt{(l+1+j)(N-l-2j)} \\
        &\hspace{1.1cm}+\lambda_1(N,l,j+1) \sqrt{(N-l-1-2j)(j+1)} \Big] \\
        &\hspace{1.1cm}\ket{l+1+j,N-l-1-2j,j}_F.
    \end{split}
\end{equation}
Since every summand has to vanish, the latter translates into a recurrence relation for the coefficients, namely
\begin{equation}
    \hspace{-0.1cm}\lambda_1 (N,l,j+1) = -\sqrt{\frac{l+j+1}{j+1} \frac{N-l-2j}{N-l-2j-1}} \lambda_1 (N,l,j).
\end{equation}
Using elementary properties of factorials and double factorials, we can immediately read off its solution
\begin{equation}
    \lambda_1 (N,l,j)=\frac{(-1)^j}{c} \sqrt{\frac{(l+j)!}{j!}\frac{(N-l-2j-1)!!}{(N-l-2j)!!}},
\end{equation}
where $c = \lambda_1 (N,l,0)$ is a normalization constant. We obtain its value by the state's normalization, which requires
\begin{equation}
    \begin{split}
        \hspace{-0.1cm}c^2 &= \sum_{j = 0}^{\frac{N-l}{2}} \frac{(l+j)!}{i!}\frac{(N-l-2j-1)!!}{(N-l-2j)!!} \\
        &= \frac{l! \, \Gamma\left(\frac{N-l+1}{2}\right)}{\sqrt{\pi} \left( \frac{N-l}{2} \right)!}\,_2F_1\left(\frac{l-N}{2},l+1;\frac{l-N+1}{2};1\right) \\
        &= (-1)^{\frac{N-l}{2}} \frac{l!}{\left( \frac{N-l}{2} \right)!} \frac{\Gamma (- \frac{1}{2} - l)}{\Gamma (- \frac{N+l+1}{2})},
    \end{split}
\end{equation}
where we used Gauss' hypergeometric theorem to simplify the Gaussian hypergeometric function $_2F_1$. Putting everything together leads to
\begin{equation}
    \begin{split}
        \lambda_1 (N,l,j) &= \frac{(-1)^j (-i)^{\frac{N-l}{2}}}{\sqrt[4]{\pi}} \sqrt{\left( \frac{N-l}{2} \right)! \binom{l+j}{j} } \\
        &\hspace{0.4cm} \sqrt{\frac{\Gamma (\frac{N-l+1}{2}-1)}{\Gamma (\frac{N-l-2}{2}-j)} \frac{\Gamma (-\frac{N+l+1}{2})}{\Gamma (-\frac{1}{2}-l)} }.
    \end{split}
    \label{eq:ThreeModeCoefficients1}
\end{equation}

\section{Integer powers of the three-mode lowering operator}
\label{app:ThreeModeLoweringOperator}
We define the composite operators
\begin{equation}
    \boldsymbol{p} = \boldsymbol{a}_1 \boldsymbol{a}_2^{\dagger}, \quad \boldsymbol{q} = \boldsymbol{a}_2 \boldsymbol{a}_3^{\dagger}, \quad \boldsymbol{r} \equiv [\boldsymbol{q}, \boldsymbol{p}] = \boldsymbol{a}_1 \boldsymbol{a}_3^\dagger,
\end{equation}
such that $\boldsymbol{L}_- = \boldsymbol{p} + \boldsymbol{q}$, with commutation relations
\begin{equation}
     [\boldsymbol{p}, \boldsymbol{r}] = [\boldsymbol{q}, \boldsymbol{r}] = 0, \quad [\boldsymbol{q}, \boldsymbol{p}^d] = d \boldsymbol{r} \boldsymbol{p}^{d-1}.
\end{equation}
Using these definitions, we rewrite \eqref{eq:ThreeModeLoweringOperatorPowers} as
\begin{equation}
    \boldsymbol{L}_-^d = \sum_{e=0}^{\floor{\frac{d}{2}}} \sum_{f=0}^{d-2e} \lambda_3 (d,e,f)  \boldsymbol{p}^{d-2e-f} \boldsymbol{q}^f \boldsymbol{r}^e.
    \label{eq:ThreeModeLoweringOperatorPowersClaim}
\end{equation}
We prove the latter by induction over $d$. For the base case $d=1$, the hypothesis is true since
\begin{equation}
    \boldsymbol{L}_- = \lambda_3 (1,0,0) \boldsymbol{p} + \lambda_3 (1,0,1) \boldsymbol{q} = \boldsymbol{a}_1 \boldsymbol{a}_2^\dagger + \boldsymbol{a}_2 \boldsymbol{a}_3^\dagger.
\end{equation}
Now, assuming that \eqref{eq:ThreeModeLoweringOperatorPowersClaim} holds for some $d \in \mathbb{N}$, we compute the $(d+1)$th. term, to wit
\begin{equation}
    \begin{split}
        \boldsymbol{L}^{d+1}_- &= \boldsymbol{L}_- \, \boldsymbol{L}^{d}_- \\
        &= \sum_{e=0}^{\floor{\frac{d}{2}}} \sum_{f=0}^{d-2e} \lambda_3(d,e,f) \Big[\boldsymbol{p}^{(d+1)-2e-f} \boldsymbol{q}^f \boldsymbol{r}^e \\
        &\hspace{1cm}+ \boldsymbol{p}^{d-2e-f} \boldsymbol{q}^{f+1} \boldsymbol{r}^e \\
        &\hspace{1cm}+ (d-2e-f) \boldsymbol{p}^{d-1-2e-f} \boldsymbol{q}^f \boldsymbol{r}^{e+1}],
    \end{split}
    \label{eq:ThreeModeLoweringOperatorInductionStep1}
\end{equation}
where we employed the identity $\boldsymbol{q} \boldsymbol{p}^{d} = \boldsymbol{p}^{d} \boldsymbol{q} + d \boldsymbol{r} \boldsymbol{p}^{d+1}$. 

For the first summand in the latter expression, the sum over $f$ can be extended to $(d+1)-2e$ since $\lambda_3 (d,e,d+1-2e)=0$. For the second term, we first shift the second running index $f \to f+1$, and then extend the sum's lower range from $f=1$ to $f=0$ since $T(d,e,-1)=0$. For the third term, we instead shift the first running index $e \to e+1$ and also extend the sum to $e=0$ as $\lambda_3 (d,-1,f)=0$. Thereafter, we restrict this term's upper limit of the sum over $f$ to $(d+1)-2e$ as the coefficient $(d+2)-2e-f$ vanishes when $f=(d+2)-2e$. In summary, the latter expression simplifies to
\begin{equation}
    \begin{split}
        \boldsymbol{L}_-^{d+1} &= \sum_{e=0}^{\floor{\frac{d}{2}}} \sum_{f=0}^{(d+1)-2e} \Big[ \lambda_3 (d,e,f) + \lambda_3 (d,e,f-1) \\
        &\hspace{1cm}+ (d+2-2e-f) \lambda_3 (d,e-1,f) \Big] \\
        &\hspace{1cm}\times\boldsymbol{p}^{(d+1)-2e-f} \boldsymbol{q}^f \boldsymbol{r}^e + \boldsymbol{Z} (d),
    \end{split}
\end{equation}
where the last term 
\begin{equation}
    \boldsymbol{Z} (d) = \begin{cases}
        0 & d \text{ even}, \\
        \lambda_3 (d,\frac{d-1}{2},0) \boldsymbol{r}^{\frac{d+1}{2}} & d \text{ odd},
    \end{cases}
\end{equation}
corresponds to the upper limit $\floor{d/2}+1$ of the sum over $e$ in the third term of \eqref{eq:ThreeModeLoweringOperatorInductionStep1} after shifting $e \to e + 1$. Note here that $\lambda_3 (d,e,f) = 0$ if $e$ or $f$ is negative, which is consistent with viewing the factorial as a Gamma function in a limiting sense. The coefficient in the remaining sum simplifies to
\begin{equation}
    \begin{split}
        \boldsymbol{L}_-^{d+1} &= \sum_{e=0}^{\floor{\frac{d}{2}}} \sum_{f=0}^{(d+1)-2e} \lambda_3(d+1,e,f) \boldsymbol{p}^{(d+1)-2e-f} \boldsymbol{q}^f \boldsymbol{r}^e \\
        &\hspace{0.4cm}+ \boldsymbol{Z} (d).
    \end{split}
\end{equation}

At last, we distinguish between even and odd $d$. In the former case, we find $\floor{(d+1)/2} = d/2 = \floor{d/2}$ and $\boldsymbol{Z}(d) = 0$, which immediately proves \eqref{eq:ThreeModeLoweringOperatorPowersClaim} for even $d$. In the latter case, we have 
$\floor{(d+1)/2} = (d+1)/2 = \floor{d/2}+1$ and thus 
\begin{equation}
    \begin{split}
        \boldsymbol{L}_-^{d+1} &= \sum_{e=0}^{\frac{d+1}{2}} \sum_{f=0}^{(d+1)-2e} \lambda_3(d+1,e,f) \boldsymbol{p}^{(d+1)-2e-f} \boldsymbol{q}^f \boldsymbol{r}^e \\
        &\hspace{0.4cm}-\left[\lambda_3 \left(d+1,\frac{d+1}{2},0 \right) - \lambda_3 \left(d,\frac{d-1}{2},0\right) \right] \boldsymbol{r}^{\frac{d+1}{2}}.
    \end{split}
\end{equation}
Since $\lambda_3 (d+1,\frac{d+1}{2},0) = \lambda_3 (d,\frac{d-1}{2},0)$, the last term vanishes, thereby proving \eqref{eq:ThreeModeLoweringOperatorPowersClaim} also for odd $d$.

\section{Coefficients of arbitrary three-mode spin states}
\label{app:ThreeModeCoefficients2}
We first recall the identities
\begin{equation}
    \begin{split}
        \boldsymbol{a}^y \ket{x} &= \sqrt{\frac{x!}{(x-y)!}} \Theta (x-y) \ket{x-y}, \\
        (\boldsymbol{a}^\dagger)^y \ket{x} &= \sqrt{\frac{(x+y)!}{x!}} \ket{x+y},
    \end{split}
\end{equation}
where $\Theta (x)$ denotes the Heaviside $\Theta$-function with the convention $\Theta (0)=1$ understood. When applying \eqref{eq:ThreeModeLoweringOperatorPowers} to \eqref{eq:ThreeModeHighestWeightState}, it is straightforward to verify that \\
\begin{widetext}
    \begin{equation}
        \begin{split}
            \lambda_4 (N,l,l_z,j,e,f) &= \sqrt{\frac{(l+j)!}{(l_z+j+e+f)!} \frac{(N-l_z-2j-2e-2f)!}{(N-l-2j-f)!} \frac{(N-l-2j)!}{(N-l-2j-f)!} \frac{(j+e+f)!}{j!}} \\
            &\hspace{0.4cm} \times \Theta (l_z + j + e + f) \Theta (N - l -2j - f)
        \end{split}.
    \end{equation}  
\end{widetext}    
This together with \eqref{eq:ThreeModeCoefficients1}, \eqref{eq:ThreeModeCoefficients2} and \eqref{eq:ThreeModeCoefficients3} shows that the full coefficient in \eqref{eq:ThreeModeSpinState} reads
\begin{widetext}
    \begin{equation}
        \begin{split}
            \lambda (N,l,l_z,j,e,f) &= \lambda_1 (N,l,j) \times \lambda_2  (l,l_z) \times \lambda_3 (l-l_z,e,f) \times \lambda_4 (N,l,l_z,j,e,f)\\
            &= \frac{(-1)^j (-i)^{\frac{N-l}{2}}}{\sqrt[4]{\pi}} \sqrt{\left( \frac{N-l}{2} \right)! \binom{l+j}{j} } \sqrt{\frac{\Gamma (\frac{N-l+1}{2}-1)}{\Gamma (\frac{N-l-2}{2}-j)} \frac{\Gamma (-\frac{N+l+1}{2})}{\Gamma (-\frac{1}{2}-l)} } \\
            &\hspace{0.4cm}\times \sqrt{\frac{2^{l-l_z} (l+l_z)!}{(2l)!(l-l_z)!}} \times \frac{(l-l_z)!}{2^e e! f! (l-l_z-2e-f)!} \\
            &\hspace{0.4cm}\times \frac{1}{(N-l-2j-f)!} \sqrt{\frac{(l+j)! (N-l_z-2j-2e-2f)! (N-l-2j)! (j+e+f)!}{j! (l_z+j+e+f)!}} \\
            &\hspace{0.4cm} \times \Theta (l_z + j + e + f) \Theta (N - l -2j - f).
        \end{split}
    \end{equation}
\end{widetext}

\section{Coefficient of W state}
\label{app:ThreeModeCoefficientsGHZW}
The W state's coefficient follows from the overlap 
\begin{equation}
    \lambda_{\text{W}} (N,l) = \,_F\braket{0,N,0|N,l,0}_s.
\end{equation}
We calculate the latter using \eqref{eq:ThreeModeSpinState}. The only contribution to the threefold sum comes from the term for which $j=e=f=0$ since
\begin{equation}
    \begin{split}
        &\,_F\braket{0,N,0|j+e+f,N-2j-2e-2f,j+e+f}_F \\
        &= \delta_{j+e+f \, 0},
    \end{split}
\end{equation}
and $j,e,f \ge 0$. Hence,
\begin{equation}
    \lambda_{\text{W}} (N,l) = \lambda (N,l,0,0,0,0).
\end{equation}

For the special case where $l=N$, we can verify the latter also by nothing that
\begin{equation}
    \begin{split}
        &\,_F\braket{0,N+1,0 | N+1,N+1,0}_s \\
        &= \sqrt{\frac{2^{N+1}}{(2N+2)! (N+1)}} \\
        &\hspace{0.4cm} \,_F\braket{0,N,0|\boldsymbol{a}_2 \boldsymbol{L}_-^{N+1}|N+1,N+1,N+1}_s.
    \end{split}
\end{equation}
Since the spin state on the right-hand-side is the highest weight state and $[\boldsymbol{a}_2, \boldsymbol{L}_-^{N+1}] = (N+1) \boldsymbol{L}_-^{N} \boldsymbol{a}_1$, this overlap simplifies to
\begin{equation}
    \begin{split}
        &\,_F\braket{0,N+1,0 | N+1,N+1,0}_s \\
        &=\sqrt{\frac{N+1}{2N+1}} \,_F\braket{0,N,0 | N,N,0}_s.
    \end{split}
\end{equation}
Hence, by recursion
\begin{equation}
    \,_F\braket{0,N,0 | N,N,0}_s = \sqrt{\frac{N!}{(2N-1)!!}},
\end{equation}
in agreement with \eqref{eq:ThreeModeCoefficientsGHZW}.

\textcolor{white}{.}
\vspace{2cm}
\clearpage


\bibliography{references.bib}

\end{document}